\documentclass[12pt]{iopart}
\pdfoutput=1
\expandafter\let\csname equation*\endcsname\relax
\expandafter\let\csname endequation*\endcsname\relax
\usepackage[numbers,sort&compress]{natbib} 
\usepackage[pdftex]{graphicx,color}
\usepackage{amsmath,amsfonts,amssymb,wasysym,graphicx,capt-of,ifthen,calc}
\usepackage{enumerate,float,url,color,subfigure}

\begin{document}
\title{Polarization and Consensus by Opposing External Sources}
\author{Deepak Bhat}
\address{Santa Fe Institute, 1399 Hyde Park Rd., Santa Fe, New Mexico 87501, USA}
\author{S. Redner}
\address{Santa Fe Institute, 1399 Hyde Park Rd., Santa Fe, New Mexico 87501, USA}

\begin{abstract}

  We introduce a socially motivated extension of the voter model in which
  individual voters are also influenced by two opposing, fixed-opinion news
  sources.  These sources forestall consensus and instead drive the
  population to a politically polarized state, with roughly half the
  population in each opinion state.  Two types social networks for the voters
  are studied: (a) the complete graph of $N$ voters and, more realistically,
  (b) the two-clique graph with $N$ voters in each clique.  For the complete
  graph, many dynamical properties are soluble within an annealed-link
  approximation, in which a link between a news source and a voter is
  replaced by an average link density.  In this approximation, we show that
  the average consensus time grows as $N^\alpha$, with
  $\alpha = p\ell/(1-p)$.  Here $p$ is the probability that a voter consults
  a news source rather than a neighboring voter, and $\ell$ is the link
  density between a news source and voters, so that $\alpha$ can be greater
  than 1.  The polarization time, namely, the time to reach a politically
  polarized state from an initial strong majority state, is typically much
  less than the consensus time.  For voters on the two-clique graph, either
  reducing the density of interclique links or enhancing the influence of
  news sources again promotes polarization.

\end{abstract}

\maketitle 

 
\section{Introduction}

News sources play a pivotal role in influencing public opinion, and the
manner by which they influence society is complex.  Each of us is bombarded
with often conflicting narratives that originate from news sources with
different viewpoints.  Some news sources are authoritative and others are
trivial and/or wrong.  In such a cacophonous environment, how does public
opinion change in time?  Motivated by this basic question, we introduce a
simple extension of the classic voter model
(VM)~\cite{C73,HL75,Cox,L85,K92,DCCH01,CFL09,KRB10,B18} to investigate how
opposing news sources influence the opinions of individuals.

The VM provides an idealized description of the opinion dynamics in a
population that consists of $N$ voters, each of which can be in one of two
possible opinion states, denoted as $+$ and $-$.  In the VM the opinion of
each voter changes in an elemental update event as follows: a randomly
selected voter adopts the opinion of a randomly selected neighbor.  This
updating is repeated until a finite population necessarily reaches consensus.

Two well-known and basic characteristics of the VM are: (a) the exit
probability and (b) the consensus time.  The exit probability is defined as
the probability for the population to reach + consensus as a function of the
initial fraction $x$ of + voters.  The consensus time is defined as the
average time for the population to reach consensus (either $+$ or $-$) as a
function of $N$ and $x$.  The dependences of the exit probability and the
consensus time on $N$ and $x$ have been fully characterized for a wide range
of underlying networks~\cite{SEM04,SEM05,CLB05,SR05,ARS06,SAR08,VE08}.

While the VM is compelling because of its simplicity and natural
applications, the model is much too naive to account for
opinion formation of a real society.  A wide variety of extensions of the
VM have therefore been proposed that incorporate more realistic
features of individual opinion changes.  Some examples include:
zealotry~\cite{M03,MPR07,GJ07}, where some voters never change opinion,
adaptation~\cite{GDB06,HN06,KB08,NKB08,SS08,SS10,DGL12,RG13}, where the underlying
network connections change in response to opinion changes,
vacillation~\cite{LR07}, where a voter may consult multiple neighbors before
changing opinion, latency~\cite{LSB09}, where a voter must ``wait'' after an
opinion change before changing again, heterogeneity~\cite{MGR10}, where each
voter has a distinct rate to change opinion, and reputation~\cite{Bhat},
where a dynamically changing individual reputation determines how likely a
voter can influence the opinion of a neighbor.  Some of these extensions are
discussed in a recent review~\cite{R19}.

While much rich phenomenology has been uncovered by these studies, consensus
is not the typical outcome for many decision-making processes.  This basic
fact has motivated additional extensions of the VM in which consensus can be
forestalled as a natural outcome of the dynamics.  Some examples include:
stochastic noise~\cite{Manfred,Boris,Adrian}, the influence of multiple
neighbors~\cite{Castellano-q}, self confidence~\cite{Volovik},
partisanship~\cite{XSK11,Masuda2}, and multiple opinion
states~\cite{deffuant2002can,hegselmann2002opinion,ben2003bifurcations}.

Within the rubric of hindering consensus, a natural mechanism is the
influence of external and competing news sources.  In this work, we introduce
a simple extension of the VM in which voters are influenced both by their
neighbors and by two news sources with fixed and different opinions.  Each
news source is connected to a specified subset of voters, which may be
disjoint or overlapping.  A news source can influence individual voters but
the news sources are not influenced by public opinion.  Our goal is to
characterize when the population reaches consensus and when it is driven to a
politically polarized state, with roughly half of the voters in each voting
state, as a function of the persuasiveness of the news sources.

In Sec.~\ref{formalism}, we briefly outline the theoretical approaches that
will be used to quantify the properties of our model.  We will focus on the
exit probability, the consensus time, and the \emph{polarization time},
namely, the time for the population to reach a politically polarized state of
50\% $+$ voters and 50\% $-$ voters when starting from a state with unequal
densities of $+$ and $-$ voters.  We then discuss the basic properties of the
model when voters reside on a complete graph with two opposing news sources
(Sec.~\ref{completegraph}).  In Sec.~\ref{twocliquegraph}, we treat the model
in the more realistic situation where voters reside on a two-clique graph
with each news source linked to only one of the cliques.  We give a brief
summary In Sec.~\ref{conclusions}.

\section{Formalism}
\label{formalism}

We first introduce the basic quantities that will be studied in this work.
We denote by $x$ the fraction of voters with $+$ opinion at any time $t$, and
$y$ as the initial fraction of $+$ voters at $t=0$.  We define the exit
probability $E_+(y)$ as the probability that a population of $N$ voters
reaches $+$ consensus when the initial fraction of $+$ voters is $y$.
Correspondingly, $E_-(y)=1-E_+(y)$ is the probability for the population to
reach $-$ consensus from the same initial state.  The consensus time
$T_{\rm con}(y)$ is defined as the average time for a population of $N$
voters to reach $+$ or $-$ unanimity when the initial fraction of $+$ voters
equals $y$.  We are typically interested in the initial condition
$y=\frac{1}{2}$; in this case, we write the consensus time as $T_{\rm con}$,
with no argument.

In the presence of opposing news sources, there exists another characteristic
and distinct time scale that we term the \emph{polarization time},
$T_{\rm pol}(y)$.  This quantity is defined as the average time for the
population to reach the politically polarized state, with equal densities of
$+$ and $-$ voters, when the initial fraction of $+$ voters equals $y$, which
we take as less than $\frac{1}{2}$ without loss of generality.  The
polarization time quantifies the effectiveness of the opposing news sources
to promote their viewpoints and thereby forestall the consensus that would
arise if individuals only interacted amongst their peers.  A natural initial
condition for the polarization time is $y=0$; that is, starting from $-$
consensus.  This state is not a fixed point of the stochastic dynamics
because of the presence of the $+$ news source that pulls the population away
from $-$ consensus whenever this state is reached.  For this initial
condition, we write the polarization time as $T_{\rm pol}$, again with no
argument.

The time evolution of opinions is controlled by the rates for $x$ to
change by $\pm \frac{1}{N}\equiv \pm\delta x$ in a single update event; these
are defined as $r^{\pm}(x)$ respectively.  In terms of these microscopic
rates, the probability $P(x,t)\,\delta x$ that the fraction of $+$ voters
lies between $x$ and $x+\delta x$ changes in time according to the master
equation
\begin{subequations}
\begin{align}
  \label{ME}
  \frac{\partial P}{\partial t}=r^+(x\!-\!\delta x) P(x\!-\!\delta x,t)
                         +r^-(x\!+\!\delta x) P(x\!+\!\delta x,t)
-\left[r^+(x)\!+\!r^-(x)\right]P(x,t)\,.
\end{align}
Expanding this equation in a Taylor series to second order gives the
Fokker-Planck equation
\begin{align}
 \label{FP}
  \frac{\partial P}{\partial t} =-\frac{\partial }{\partial x}[V(x)P]
  + \frac{\partial^2 }{\partial x^2}[D(x)P]\,,
\end{align}
\end{subequations}
where the drift velocity and diffusion coefficient are 
\begin{align}
\begin{split}
\label{VD}    
  V(x)&=\big[r^+(x)-r^-(x)\big] \delta x\,,\\ 
  D(x)&=\big[r^+(x)+r^-(x)\big] (\delta x^2/2)\,,
\end{split}  
\end{align}
respectively.  We can view the instantaneous opinion $x$ as undergoing biased
and position-dependent diffusion in the interval $[0,1]$ in the presence of
the effective potential
\begin{align}
  \label{potential}
\phi(x)=-\int^{x} \frac{V(x')}{D(x')}\,dx'\,.
\end{align}
As we shall see, the nature of this potential determines the $N$ dependence of
the consensus and polarization times.

To determine the exit probability, as well as the consensus and polarization
times, we use the backward equation approach, a basic tool of first-passage
processes~\cite{G85,K97,R01}.  This approach relies on the fact that the
opinion state of the population ``renews'' itself after each microscopic
update event.  In this framework, the exit probability satisfies the backward
equation
\begin{subequations}
\begin{align}
  \label{Ey}
  E_+(y)= \varepsilon E_+(y+\delta y)+ (1-\varepsilon) E_+(y-\delta y) \,, 
\end{align}
where $\varepsilon=r^+(y)/[r^+(y)\!+\!r^-(y)]$ is the probability for $y$ to
increase by $\delta y=\frac{1}{N}$ in a single update.  This equation merely
states that the exit probability starting from the state $y$ is a weighted
average of the exit probabilities after one update step.  Namely, with
probability $\varepsilon$, $y\to y+\delta y$, at which point the exit
probability is $E_+(y+\delta y)$.  Conversely, with probability
$1-\varepsilon$, $y\to y-\delta y$, at which point the exit probability is
$E_+(y-\delta y)$.  Expanding \eqref{Ey} in a Taylor series to second order
gives
\begin{align}
  \label{ldagger-E}
 V(y) \frac{\partial E_+}{\partial y}+ D(y) \frac{\partial^2 E_+}{\partial y^2}=0\,. 
\end{align}
\end{subequations}
This equation is subject to the boundary conditions $E_+(0)=0$, $E_+(1)=1$;
that is, when $y=1$, exit to the state $y=1$ occurs with probability 1, while
when $y=0$, exit cannot occur.  The formal solution is
\begin{align}
 \label{eqE}
 E_+(y)=  \frac{{ \int^{y}_{0} \exp[\phi(y')]dy'}}
  {{ \int_0^1 \exp[\phi(y')]dy'}}\,,
\end{align} 
and normalization gives $E_-(y)=1-E_+(y)$.

Using this same reasoning, the consensus and polarization times satisfy the
backward equation~\cite{G85,K97,R01},
\begin{subequations}
\begin{align}
  \label{backward}
T(y)= \varepsilon[T(y+\delta x)+dt]+(1-\varepsilon)[T(y-\delta y)+dt] \,.
\end{align} 
Here $dt=[r^+(y)+r^-(y)]^{-1}$ is the time for an elemental update from the
state $y$.  Expanding Eq.~\eqref{backward} in a Taylor series to second order
now gives
\begin{align}
   \label{ldagger-T}
 V(y) \frac{\partial T}{\partial y}+ D(y) \frac{\partial^2 T}{\partial y^2}=-1\,,
\end{align}
\end{subequations}
with distinct boundary conditions for the consensus and polarization times.
For the consensus time, the boundary conditions are $T(0)=T(1)=0$; that is,
the consensus time starting from either consensus state is zero.  For the
polarization time, the appropriate boundary conditions are $T(\frac{1}{2})=0$
and $\frac{\partial T}{\partial y}\big|_{y=0}=0$.  That is, starting from the
polarized state, the polarization time is zero, while the polarization time
obeys the no flux condition if consensus is reached.  Then latter boundary
condition arises because the consensus state is not an attractor of the
stochastic dynamics.  If consensus happens to be reached, the two opposing
fixed-opinion news sources pull the population away from consensus.

The formal solutions for the consensus and polarization times are
  \begin{align}
  \begin{split}
    \label{TcTp}
     T_{\rm con}(y)&=E_+(y)I(y,1)- E_-(y)I(0,y)\, ,  \\[1mm]
   T_{\rm pol}(y)&=I(y,1/2)\, ,
  \end{split}
    \end{align}
where 
\begin{align*}
  I(a,b)=\int^{b}_{a}dy'\int^{y'}_{0}dy''\,\frac{\exp[\phi(y')-\phi(y'')]}{D(y'')}~. 
\end{align*}

In the absence of the news sources, the dynamics is simply that of the
classic VM.  When the voters reside on the complete graph, the transition
rates are (\ref{CG})
 \begin{align}
 \label{rates-vm}
r^+(x)=\tfrac{1}{2} Nx(1-x) \,,\qquad
r^-(x)=\tfrac{1}{2}Nx(1-x)\,.
\end{align}
From these rates, Eq.~\eqref{VD} gives $V(x)=0$ and $D(x)=x(1-x)/2N$, and the
full dynamics is solvable~\cite{C73,HL75,Cox,L85,K92,DCCH01,CFL09,KRB10,B18}.
Three basic results for the VM on the complete graph are: (i) $E_+(y)=y$
(which is also a consequence of magnetization conservation), (ii)
$T_{\rm con}(y)=-2N[y\ln y + (1-y) \ln (1-y)]$; that is, the consensus time
is linear in $N$, and (iii) because $x=0,1$ are natural absorbing boundaries,
$T_{\rm pol}(y)$ is infinite.  That is, when the initial fraction of $+$
voters is $y\ne \frac{1}{2}$, there is a finite probability to reach
consensus before the polarized state, which means that the polarization time
is divergent.  However, the polarization time is both finite and meaningful
when the population is also influenced by two opposing news sources.  We now
determine how the presence of news sources alters the above three properties
of the VM.
 
 \section{Voters on the complete graph}
\label{completegraph}
 
\begin{figure}[ht]
\centerline{\includegraphics[width=0.45\textwidth]{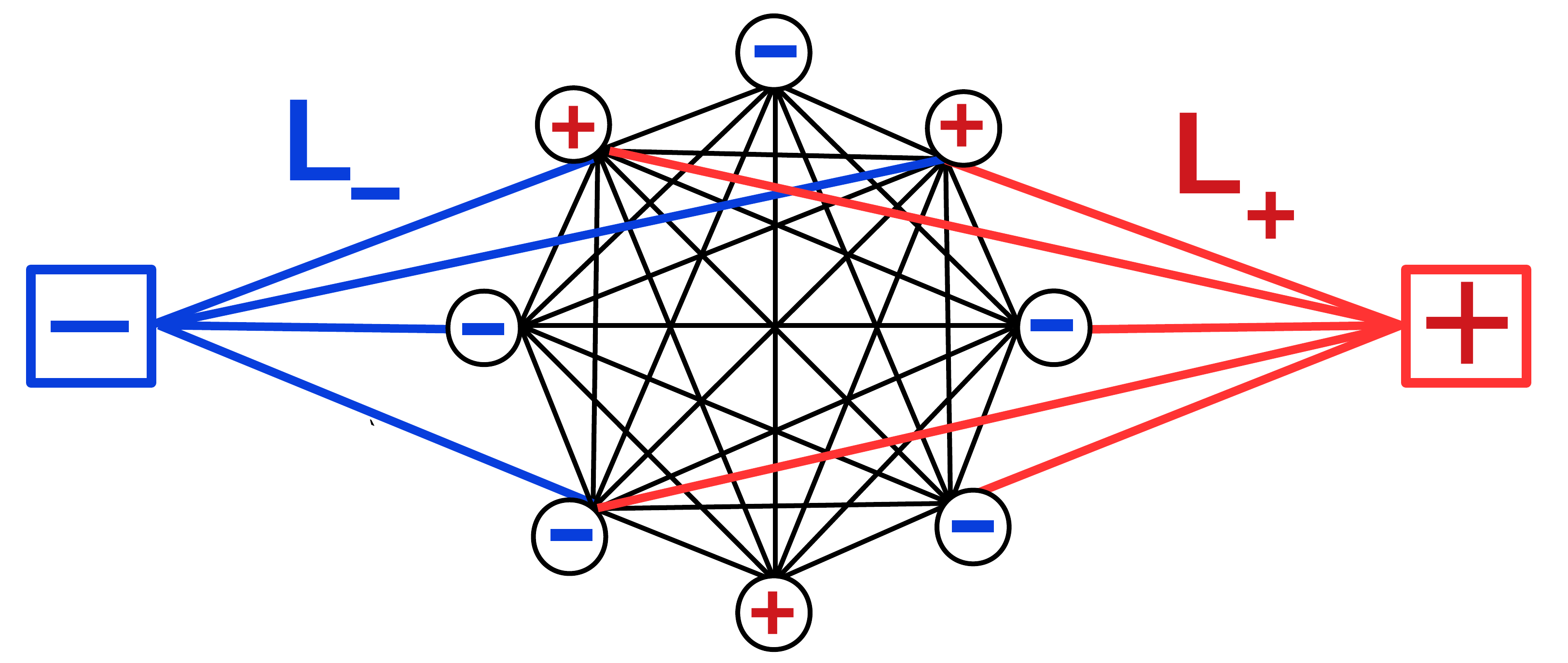}}
  \caption{ The complete-graph system.  Two opposing news sources (squares)
    influence voters (circles) on the complete graph.  The news sources have
    $L_+$ and $L_-$ links to individual voters.}
\label{fig-completegraph}
\end{figure}

Suppose that $N$ voters on the complete graph are additionally influenced by
news sources with fixed $+$ and $-$ opinions (Fig.~\ref{fig-completegraph}).
These news sources have $L_+$ and $L_-$ links to random voters, respectively,
with $0< L_{\pm}\leq N$, so that the corresponding link densities
$\ell_{\pm}=L_{\pm}/N$ lie between 0 and 1.  A basic parameter in our model
is the \emph{propensity} $p$, which quantifies the influence of a news source
on a given voter.  This propensity is implemented as follows: for a voter
that is linked to one news source and $N-1$ other voters, the news source is
picked with probability $p/R$ and a neighboring voter is picked with
probability $(1-p)(N-1)/R$, where $R=p+(N-1)(1-p)$ is the total rate of
picking any interaction partner, either neighbor or news source.  If a voter
is connected to both news sources, then $R=2p+(N-1)(1-p)$.  Once a voter has
selected an interaction partner, the voter adopts the opinion of this
partner.  This update step is repeated ad infinitum.

The opinion evolution depends on the actual connection pattern between the
news sources and voters, a situation that is analytically intractable.  This
leads us to apply a simplification that we term the \emph{annealed-link
  approximation}.  Here, we replace the true transition rates for each voter
on a given fixed-link realization by the average transition rate, in which a
link to a news source occurs with probability proportional to the appropriate
link density.  We now apply this approximation to determine the three basic
characteristics of the collective opinions, namely, $E_+(y)$,
$T_{\rm con}(y)$, and $T_{\rm pol}(y)$.  We first first need the transition
rates $r^+(x)$ and $r^-(x)$ within the annealed-link approximation.  By a
somewhat tedious but straightforward calculation (see \ref{CG2} for details),
these rates are
\begin{align}
\begin{split}
\label{eqrates}
r^+(x)&=\tfrac{1}{2}A Nx(1-x) +B_+ (1-x)\,,\\
r^-(x)&=\tfrac{1}{2}ANx(1-x)+B_-x\,.
\end{split}
\end{align}
The $A$ terms account for the rate at which voters adopt the opinion of
neighboring voters, while the $B$ terms account for opinion changes due to
the interaction of voters with news sources.  As shown in \ref{CG2}, the
amplitudes $A$ and $B_\pm$ are
\begin{subequations}
\label{AB}
\begin{align}
   \label{A}
  A=&\frac{(1\!-\!\ell_+)(1\!-\!\ell_-)}{1\!-\!(1/N)}
      +\frac{(1-p)(\ell_++\ell_--2\ell_+\ell_-)}
      {(1-p)+(2p-1)/N}
   +\frac{(1-p)\ell_+\ell_-}{(1-p)+(3p-1)/N}\,,\\[2mm]
  B_{\pm}&=\frac{p\ell_{\pm}}{2}\!
            \left[\frac{1\!-\!\ell_{\mp}}{(1\!-\!p)\!+\!(2p\!-\!1)/N}
            \!+\!\frac{\ell_{\mp}}{(1\!-\!p)\!+\!(3p\!-\!1)/N}\right]\,.
\end{align}
The three distinct terms in $A$ account for voters that are not connected to
any news source, connected to one news source, and connected to both news
sources.  Similarly, the two terms for $B_\pm$ account for voters that are
connected to one news source or to both news sources, respectively.  While
the coefficients $A$ and $B_{\pm}$ are complicated, they greatly simplify in
the large-$N$ limit, where
\begin{align}
\label{AB-simple}
  A\to 1\qquad  B_{\pm} \to   \frac{1}{2} \frac{p\ell_{\pm}}{(1-p)}~.
\end{align}
\end{subequations}

Substituting the transition rates \eqref{eqrates} in Eq.~\eqref{VD},  the
drift velocity and the diffusion coefficient are:
\begin{align}
\label{VD-complete}
    V(x)=\frac{B_+ (1-x) -B_-x}{N}\,, \qquad
    D(x)= A\,\, \frac{x(1-x)}{2N} + \frac{B_+ (1-x) +B_-x}{2N^2}~.
\end{align}
Using these quantities in the formalism of Sec.~\ref{formalism}, we can
compute the exit probability, the consensus time, and the polarization time
for different link densities $\ell_{\pm}$.  As a preliminary, we first study
the influence of a single $+$ news source on voter opinions and then turn to
the influence of two opposing news sources.

\subsection{Single news source}
\label{amassmedium}

When there is only a single news source, the opinion state of the population
is monotonically driven to consensus that is aligned with the news source.
We now determine its effectiveness in driving this consensus.

For a single news source, we set $\ell_-=0$ and $\ell_+=\ell$ in
Eq.~\eqref{VD-complete}, from which
\begin{align}
  \label{VDsingle}
  \frac{V(x)}{D(x)} =\frac{\alpha}{x + \alpha/(2N)} ~,
\end{align}
where $\alpha\equiv {2B_+}/{A}$, which approaches ${p\ell}/(1-p)$ as
$N\to\infty$.  The parameter $\alpha$ is fundamental, as it characterizes the
effectiveness of the news source in influencing the population, both by its
intrinsic persuasiveness and by the extent of its reach.

Using Eq.~\eqref{VDsingle} in \eqref{potential}, the effective potential in
which $x$ diffuses is
\begin{align}
\label{phi-single}
\phi(x)= - \alpha \ln \left(x+\frac{\alpha}{2N}\right)\,.
\end{align}
This asymmetric potential biases individual opinions towards the $+$
consensus state.  We determine the exit probability by substituting the
effective potential \eqref{phi-single} into \eqref{eqE} and performing the
integral to give
\begin{align}
\label{exit-amass}
    E_+(y)= 
 \begin{cases}
   {\displaystyle
     \frac{\left(\alpha+2Ny\right)^{1-\alpha}-\alpha^{1-\alpha}}{\left(\alpha+2N\right)^{1-\alpha}-\alpha^{1-\alpha}}}
   \qquad & \alpha\neq 1\,,\\[4.5mm]
   {\displaystyle \frac{\ln \left(2Ny+1\right)}{\ln \left(2N+1\right)}}\qquad & \alpha=1\,.
 \end{cases}
\end{align}

\begin{figure}[ht]
   \centerline{\includegraphics[width=0.45\textwidth]{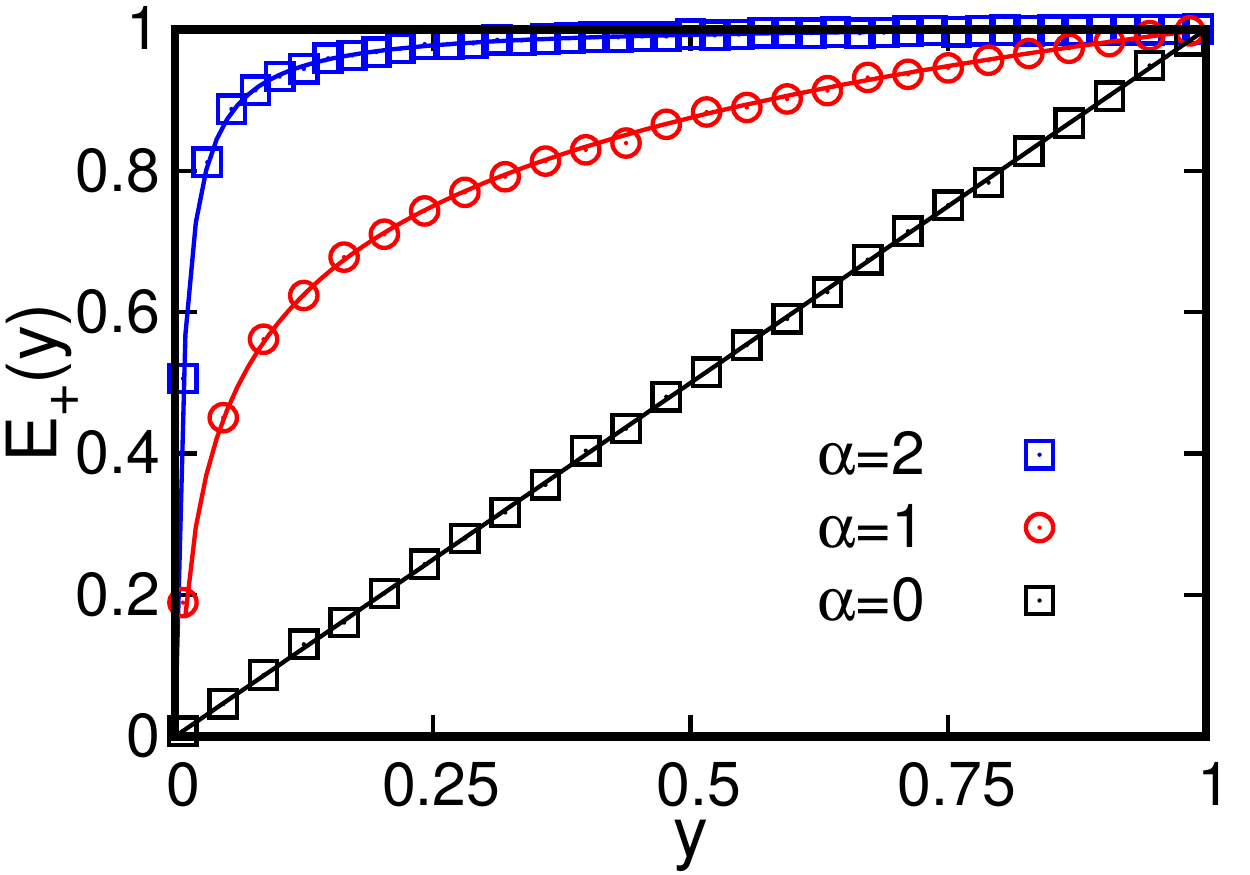}}
   \caption{\small Exit probability versus initial fraction of $+$ opinion
     voters $y$ for $N=128$ voters that are influenced by single news source
     with link density $\ell=1$.  The curves represent Eq.~\eqref{exit-amass}
     and the symbols represent simulations over $10^4$ realizations.}
\label{exit-single}  
\end{figure}

By increasing $\alpha$, the news source becomes more effective in biasing the
opinions towards $+$ consensus, as shown in Fig.~\ref{exit-single}.  Here, we
choose $(p,\ell)=(\frac{1}{2},1)$ to achieve $\alpha=1$, and
$(p,\ell)=(\frac{2}{3},1)$ to achieve $\alpha=2$.  Unless otherwise stated,
we use these parameter choices to generate systems with $\alpha=1$ and
$\alpha=2$ in subsequent figures.  The qualitative behavior in
Fig.~\ref{exit-single} is the same as exit probability of a biased random
walk on a finite interval with bias to the right~\cite{R01}.  As expected,
when the news source is very effective, the exit probability is nearly 1,
even for $y$ close to 0.

To compute the consensus time, first note, from Eq.~\eqref{VDsingle}, that
$V/D$ is of order 1, except when $x$ is of order $1/N$ or smaller.  Within
this boundary layer near $x=0$, the second term in the denominator of $V/D$
ensures that $V/D$ remains finite even when $x=0$.   We can simplify
the algebra considerably by excluding this thin boundary layer and
correspondingly dropping this second term in the denominator of $V/D$.  We
checked numerically that this approximation has a vanishingly small effect on
the consensus time for large $N$.  We determine the range of the resulting
slightly truncated interval $[a,1]$ by equating the two terms in the
denominator of $V/D$ to give $a= \alpha/(2N)$.  In this truncated interval,
we have
 \begin{align}
   \label{VD32}
   V(x)=\frac{B_+(1\!-\!x)}{N} \qquad\qquad  D(x)\approx \frac{Ax(1\!-\!x)}{2N}  \,.
\end{align}
With these simplifications, the effective potential becomes
$\phi(x)=-\alpha\ln x$ for $x \in [a,1]$.

\begin{figure}[ht]
\centerline{\subfigure[]{\includegraphics[width=0.45\textwidth]{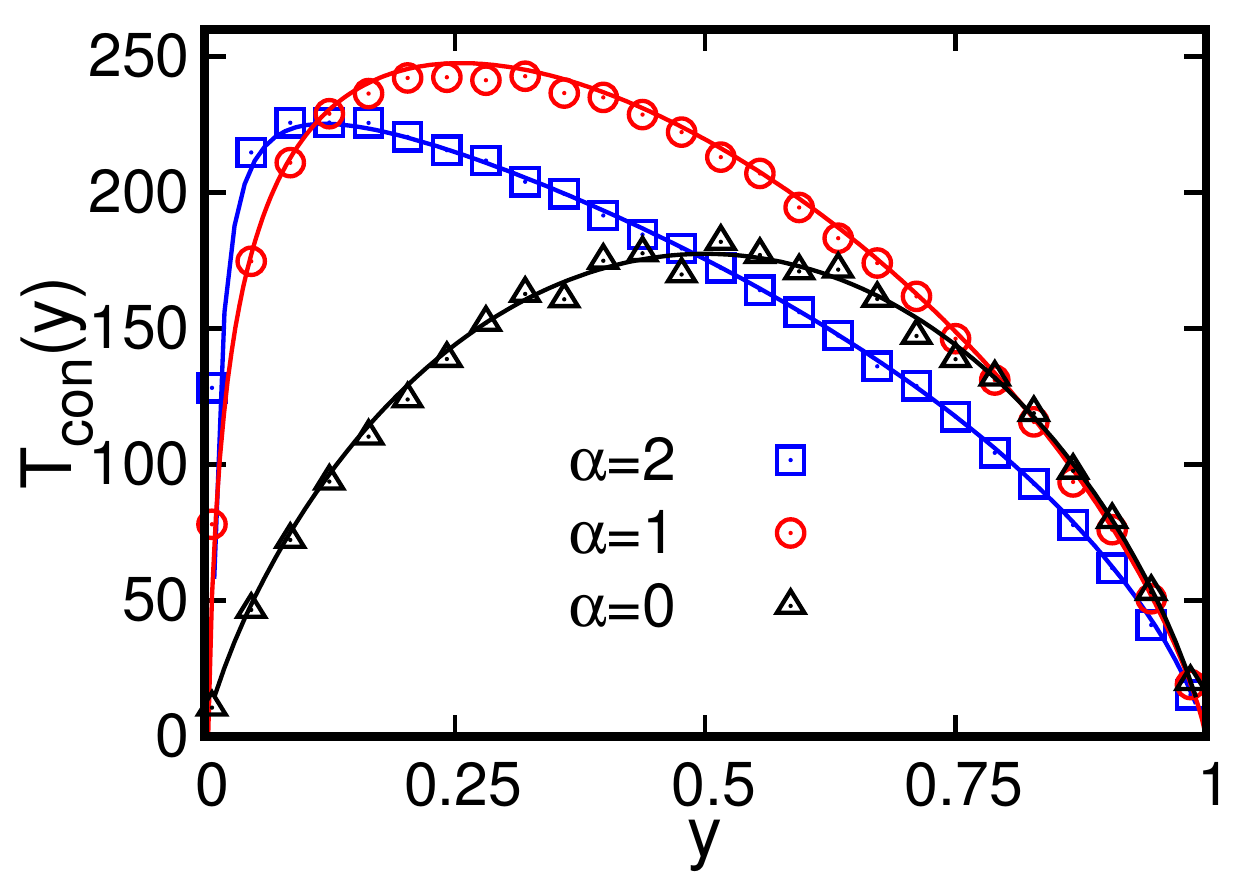}}}
\caption{\small Consensus time versus the initial fraction of $+$ opinion
  voters $y$ for $N=128$ voters with a single news source. The solid curves
  represent Eq.~\eqref{tc-amass} and the symbols represent simulation results
  over $10^4$ realizations.}
\label{tc-single}  
\end{figure}

We now substitute this effective potential, as well as the above form for
$D(x)$, into the first of Eqs.~\eqref{TcTp}.  The resulting integral can be
evaluated for certain simple values of $\alpha$.  For the cases $\alpha=1$
and $\alpha=2$, in particular, we obtain
\begin{align}
\label{tc-amass}
   T_{\rm con}(y)=
   \begin{cases}
{\displaystyle  2N\!\!\left[\frac{\ln(a/y)}{\ln a}\frac{\pi^2}{6} +\frac{ \ln y}{\ln a}\,\,{\rm  Li}_2(a)
    -{\rm Li}_2(y) \right]}   &\qquad\alpha=1\, ,\\[5.5mm]
{\displaystyle 2N\Big(1-\frac{1}{y}\Big)\ln\frac{(1-y)}{(1-a)}} &\qquad \alpha=2\, ,
\end{cases}
\end{align}
where ${\rm Li}_2(y)$ is the dilogarithm function~\cite{dilog}.  It is
possible that analytical solutions also exist for other simple values of
$\alpha$, but the results given above mostly encompass the generic behavior
for a single news source; namely, the consensus time scales linearly with
$N$, except in the limit $\alpha\to\infty$.   When the initial state is
$y=\frac{1}{2}$, we have
\begin{align*}
  T_{\rm con}=
\begin{cases}
  2N\ln 2\approx 1.386 N &\qquad \alpha=0\\
   \left(\ln^2 2+\pi^2/6\right) N\approx 2.125 N& \qquad\alpha=1\\
    2N\ln 2\approx 1.386 N &\qquad \alpha=2\,.
 \end{cases}
\end{align*}                                   
The first line is the complete-graph VM result without news sources.  Because
the news source biases the population to $+$ consensus, the maximum of
$T_{\rm con}(y)$ shifts gradually towards $x=0$, as shown in
Fig.~\ref{tc-single}.  Coincidentally, the consensus time starting from
$y=\frac{1}{2}$ is the same for $\alpha=0$ and $\alpha=2$.

In the opposite limit of $\alpha\to\infty$, the consensus time scales as
$\ln N$.  This limit is conveniently realized by choosing $\ell=1$ and
$p\to 1$.  Then Eq.~\eqref{AB} gives $A \approx \frac{1}{2} N(1-p)$ and
$B_+ \approx \frac{1}{2}N$, so that $\alpha={2B_+}/{A}\approx 2/(1-p)$.  When
$1-p \ll \frac{2}{N}$, $A$ is vanishingly small, so the forward rate
$r^+(x)\approx B_+ (1-x)$ and the backward rate $r^-(x)\approx 0$.  Thus the
fraction of $+$ voters only increases with time until $+$ consensus is
reached.  For the initial condition $x=\frac{1}{2}$, we determine the
consensus time from
\begin{align*}
  T_{\rm con} =\sum^{1-\frac{1}{N}}_{x=\frac{1}{2}} \frac{1}{r^+(x)}
  = \sum^{1-\frac{1}{N}}_{x=\frac{1}{2}} \frac{1}{B_+(1-x)} \simeq 2 \ln N\,.
\end{align*}

\subsection{Two opposing news sources}
\label{twomedia}

We now turn to our main focus of two opposing news sources.  To determine the
exit probability, as well as the consensus and polarization times, we again
need to simplify the form of the drift velocity and the diffusion coefficient
in Eq.~\eqref{VD-complete}.  Again, the ratio $V/D$ is of order 1, except
when $x$ is a distance of order $1/N$ from the boundaries at 0 and 1.  The
algebra simplifies considerably when we ignore these boundary layers.
Following the same procedure as in the previous subsection, the second term
in the denominator of $V/D$ can be neglected when $x$ is in the range
$[a_-,1-a_+]$, with $a_\pm= {B_\mp}/{AN}$.  In this truncated interval, we
may write
\begin{align}
 V(x)=\frac{\left[B_+(1\!-\!x)-B_-x\right]}{N}\qquad\qquad D(x)\approx \frac{Ax(1\!-\!x)}{2N}\,.
  \label{VD2}
\end{align}

Using this approximation for $V(x)$ and $D(x)$, the effective potential \eqref{potential} becomes
\begin{align}
  \label{phi-eff}
  \phi(x)=-\ln [x^{\alpha_+}(1-x)^{\alpha_-}]\,,
\end{align}
with $\alpha_{\pm}=2B_{\pm}/A \approx p\ell_{\pm}/(1-p)$.  Thus in the
presence of opposing news sources, the density $x$ undergoes diffusive
dynamics in the (generally) asymmetric logarithmic potential well
\eqref{phi-eff}.  Because of this well, the consensus time can be much longer
than in the case of no news sources, as we would naively expect.  However,
because the potential at the interval boundaries depends logarithmically on
$N$, the consensus time grows only algebraically, rather than exponentially,
with $N$.

\subsubsection{Symmetrically connected news sources}

For simplicity, first consider equally connected news sources and define the
common link density as $\ell \equiv \ell_{\pm}$.  Now the parameter that
quantifies the effectiveness of the news sources is
$\alpha= \alpha_{\pm} \approx p\ell/(1-p)$, while the effective potential
simplifies to $\phi(x)=- \alpha\ln \left[x(1-x)\right]$.  Moreover, the
relevant range of $x$ is $a\leq x \leq 1-a$, where
$a_+=a_-\equiv a =\alpha/(2N)$.

\begin{figure}[ht]
  \centerline{\includegraphics[width=0.45\textwidth]{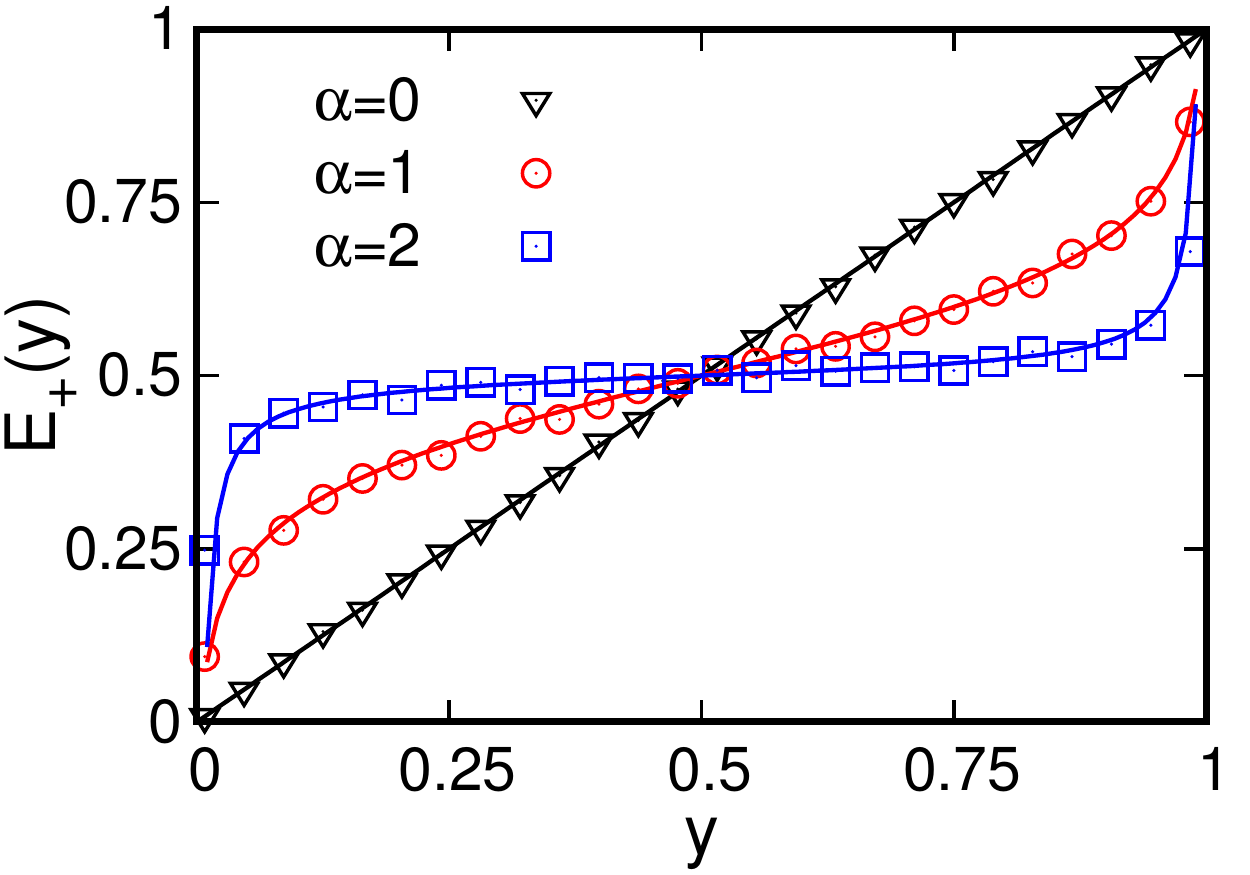}}
  \caption{\small Exit probability to $+$ consensus versus $y$ for $N=128$
    voters with opposing and symmetrically linked news sources.
    Equation~\eqref{Ep} gives the curves and the symbols correspond to
    simulations over $10^4$ realizations.}
\label{exit-double}  
\end{figure}

To obtain the exit probability, we substitute the symmetrized version of
Eq.~\eqref{VD2} into Eq.~\eqref{eqE} and evaluate the integral to obtain
\begin{align}
\label{Ep}
 E_+(y)=\frac{1}{2}\bigg[1- \frac{G_\alpha(y)}{G_\alpha(a)}\bigg]\,,
\end{align}
where, for simple rational values of $\alpha$, $G_\alpha$ can be determined
analytically.  The specific examples that we could compute are:
\begin{align*}
  G_{\frac{1}{2}}(y)&=\sin^{-1}\! \sqrt{y}-\frac{\pi}{4}\\
  G_{1}(y)&=\ln \left(y^{-1}-1\right)\\
  G_{\frac{3}{2}}(y)&=\sqrt{y(1-y)^{-1}}-\sqrt{y^{-1}(1-y)}\\
  G_2(y)&=y^{-1}-(1-y)^{-1}+\ln \left(y^{-1}-1\right)^2\,.
\end{align*}
By plotting these expressions, we find that the exit probability has an
anti-sigmoidal shape for $\alpha>0$ (Fig.~\ref{exit-double}).  This behavior
reflects the opposing role of the two news sources.  If the initial density
of the system is $y\ne\frac{1}{2}$, the news sources tend to drive the
opinions to the politically polarized state of $y=\frac{1}{2}$ before
consensus is reached.  Consequently, the exit probability becomes nearly
independent of the initial condition as the news sources become more
effective, i.e., $\alpha\gg1$.

For the consensus time, we again substitute the symmetrized form of
Eq.~\eqref{VD2} into the first of Eqs.~\eqref{TcTp} and evaluate the integral
to give
\begin{align}
\label{Tc}
 T_{\rm con}(y)=N\left[H_{\alpha}(a)-H_{\alpha}(y)\right] 
\end{align}
where, once again, $H_\alpha$ can be determined explicitly for certain 
rational values of $\alpha$:
\begin{align*}
  H_{\frac{1}{2}}(y)&=-4\sin^{-1}\!\! \sqrt{y}\,\,\, \sin^{-1}\!\! \sqrt{1-y}\,,\\
  H_{1}(y)&=-\ln \left[y(1-y)\right]\,,\\
  H_{\frac{3}{2}}(y)&=\frac{(2y-1)}{\sqrt{4y(1-y)}}
   \left[\sin^{-1}\!\! \sqrt{y}-\sin^{-1}\!\!   \sqrt{1-y}\right]\,,\\
 H_{2}(y)&=\tfrac{1}{6}\left[y^{-1}(1-y)^{-1}-2\ln \left[y(1-y)\right]\right]\,.
\end{align*}

We can determine the $N$ dependence of the consensus time from the large-$N$
behavior of the functions $H_\alpha(a)$ with $a=\alpha/(2N)$.  The dominant
contribution for large $N$ arises from the term $H_\alpha(a)$ with $a\to 0$:
\begin{align*}
  H_{\frac{1}{2}}(a)&\sim a^{1/2}\\
  H_1(a)& \sim -\ln a\\
  H_{\frac{3}{2}}(a) &\sim a^{-1/2}\\
  H_2(a)&\sim a^{-1}\,.
\end{align*}
Using $a=\alpha/2N$ and combining the above results with Eq.~\eqref{Tc}, we
find
\begin{align}
 \label{Tc-N-unequal}
T_{\rm con}  \sim 
\begin{cases}
 N &\qquad  0\leq \alpha <1\,,\\
 N\ln N &\qquad \alpha=1\,,\\
 N^{\alpha}& \qquad \alpha>1\,.
\end{cases}
\end{align}
Equation~\eqref{Tc-N-unequal} is one of our major results.

\begin{figure}[ht]
  \centerline{
    \subfigure[]{\includegraphics[width=0.32\textwidth]{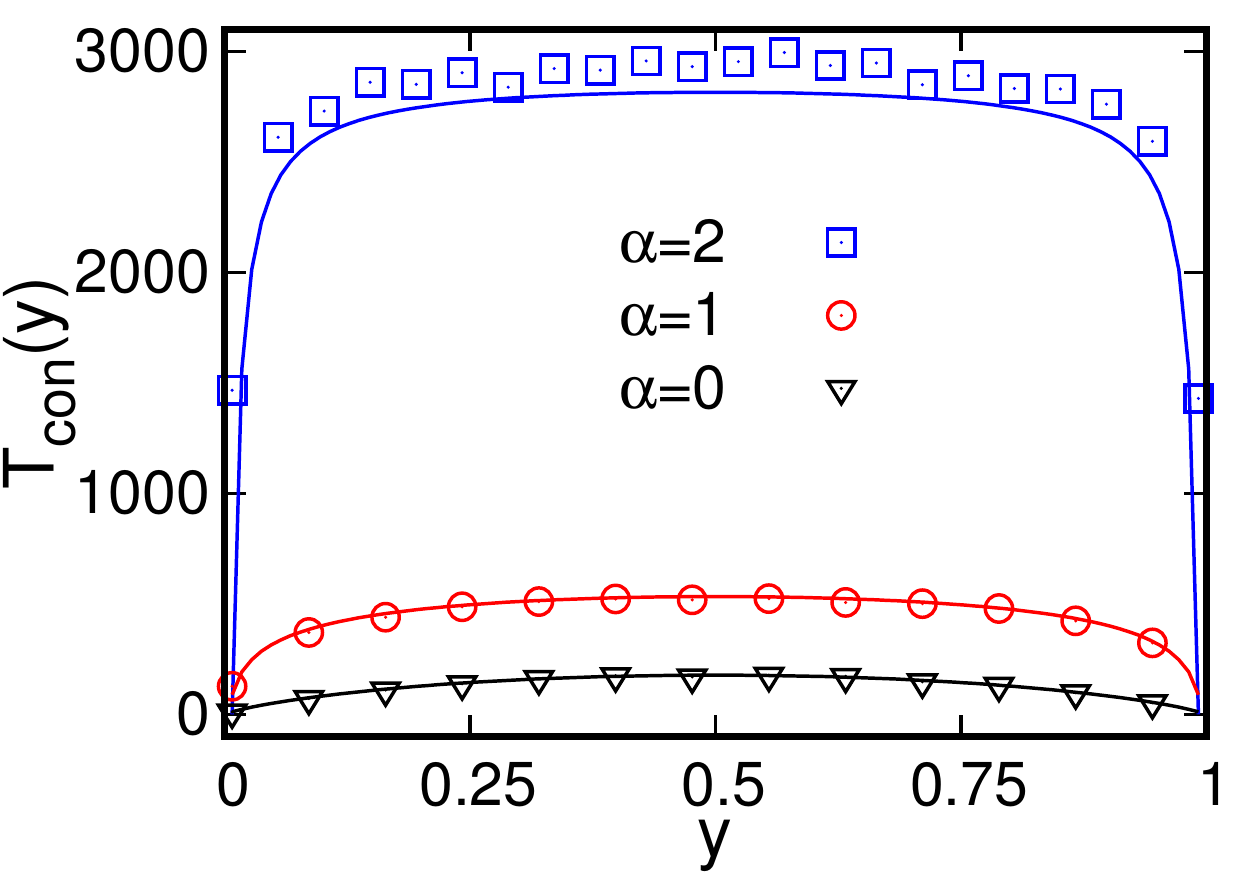}}
    \subfigure[]{\includegraphics[width=0.32\textwidth]{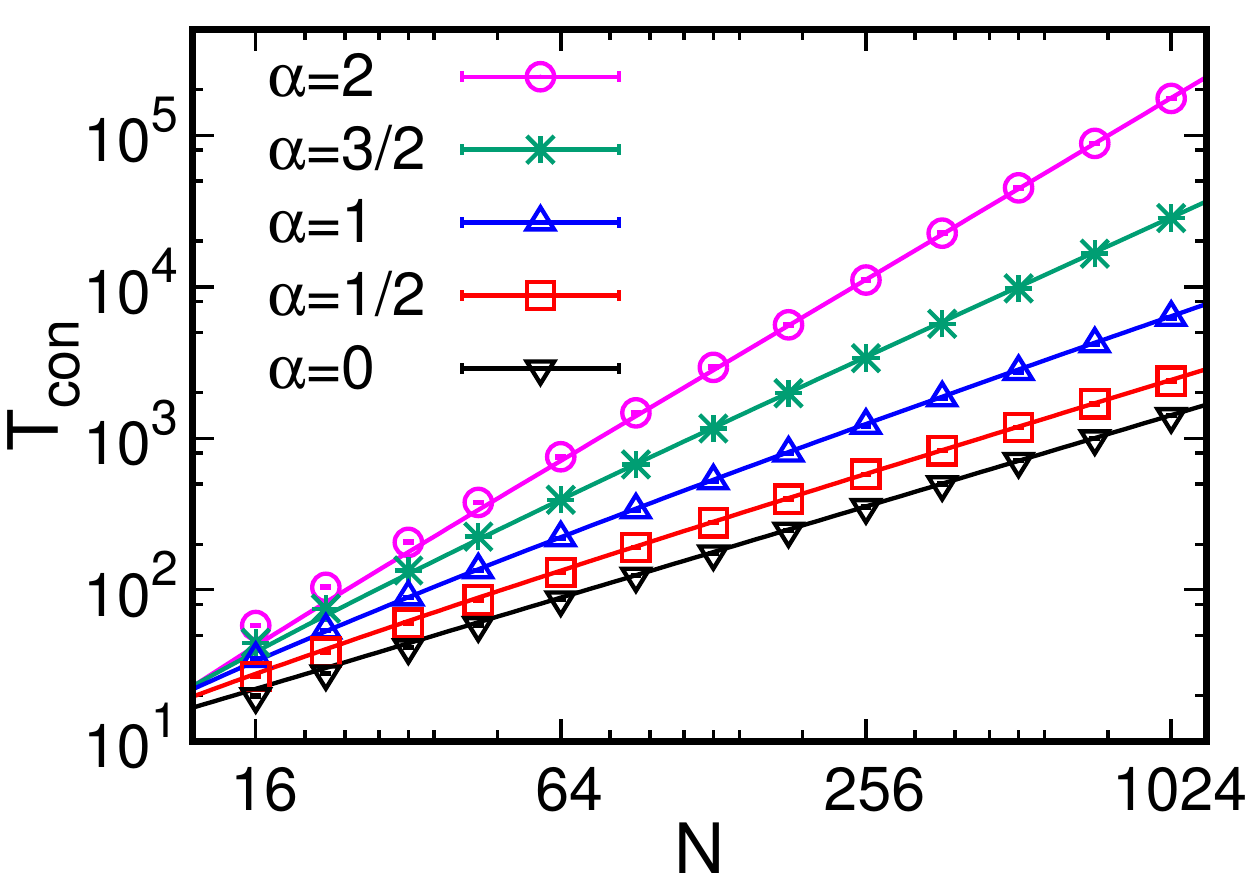}}
     {\raisebox{-3.0mm}{\subfigure[]{\includegraphics[width=0.345\textwidth]{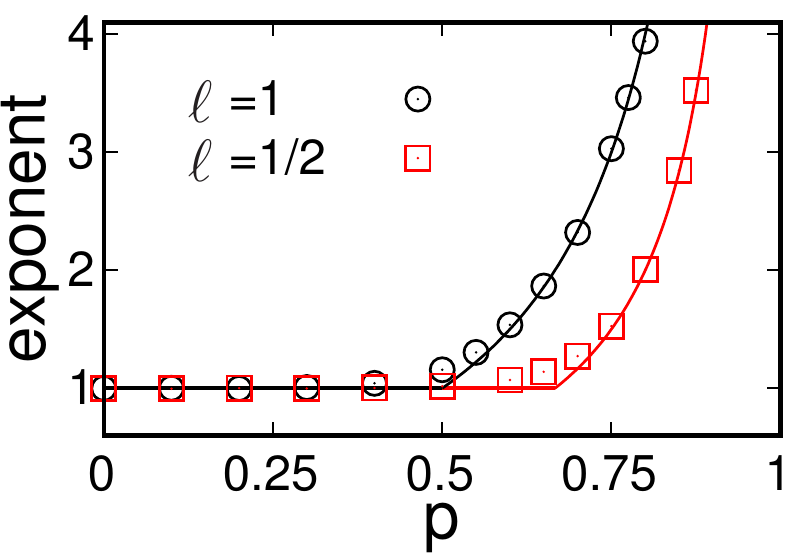}}}}}
   \caption{\small (a) Consensus time versus $y$.  Equation~\eqref{Tc} gives
     the curves and symbols represent simulations results for $N=128$ voters
     over $10^4$ realizations.  (b) Consensus time versus $N$.  The curves
     again represent the predictions from Eq.~\eqref{Tc}.  We fix $\ell=1$
     and use $\alpha=p/(1-p)$ to specify $p$ for a given $\alpha$ value.  (c)
     Consensus time exponent versus $p$ for two $\ell$ values.  The symbols
     represent simulation results and the curves are predictions from
     Eq.~\eqref{Tc-N-unequal}.}
\label{tc-double}  
\end{figure}

Our simulation results are consistent with these predictions
(Fig.~\ref{tc-double}).  A striking feature of Fig.~\ref{tc-double}(a) is
that the consensus time increases dramatically when $\alpha$ increases from 1
to 2.  This behavior reflects the different $N$ dependences of the consensus
time for $\alpha<1$ and $\alpha\geq 1$ in Eq.~\eqref{Tc-N-unequal}.  Our
estimates for the consensus time exponent for various combinations of $p$ and
$\ell$ are given in Fig.~\ref{tc-double}(c).  We determine the exponent by
extrapolating local slopes of $\ln T_{\rm con}$ versus $\ln N$ based
least-squares fits of subsets of successive data points.  For each $p$ and
$\ell$, $\alpha$ is given by $\alpha=p\ell/(1-p)$.  The sudden increase in
the exponent value at the two distinct $p$ values corresponds to the
transition at $\alpha=1$ predicted by \eqref{Tc-N-unequal}.

There are two natural ways that the news sources are connected to voters: (i)
random connections, and (ii) disjoint connections.  In the first case, a
voter may be connected to zero, one, or two news sources, while in the
latter, a voter may be connected to either zero or one news source.  For the
same link densities between the news sources and voters, we found negligible
differences in our results for the exit probability and the consensus and
polarization times.  The simulation results presented here are for the case
of random connections.

We can understand the $N$ and $\alpha$ dependences of the consensus time in a
simple way in terms of the effective potential \eqref{phi-eff}.  According
Kramers' theory~\cite{K40}, the time to reach the boundaries at $a$ and at
$1-a$ are proportional to $\exp [\phi(a)]$ and to $\exp[\phi(1-a)]$,
respectively, while the potential at these two points scales as
$\alpha \ln N$.  Consequently $T_{\rm con} \sim N^{\alpha}$ for
$\alpha>1$.  For $\alpha<1$, the effect of the logarithmic potential is
subdominant with respect to fluctuations~\cite{Hirschberg}, and it is the
latter drive the system to consensus, leading to $T_{\rm con} \sim N$.

We now determine the polarization time.  Substituting the symmetric forms of
the drift and diffusion coefficients from Eq.~\eqref{VD2} into the second of
Eq.~\eqref{TcTp} and evaluating the integral, which can be done for certain
simple values of $\alpha$, we obtain:
\begin{align}
  \label{Tp}  T_{\rm pol}(y) =
  \begin{cases}
    4N\left(\frac{\pi}{4}- \sin^{-1}\!\sqrt{y}\right)\left(\frac{\pi}{4}
  + \sin^{-1}\! \sqrt{y}-2 \sin^{-1}\! \sqrt{a}\right)   ~& \quad \alpha=\frac{1}{2}\,,\\[4mm]
 2N\left[a \ln y +(1-a) \ln (1-y)+\ln 2\right]   ~ & \quad  \alpha=1\,,\\[4mm]
 \frac{N(2y-1)}{\sqrt{y(1-y)}}\left[\sin^{-1}\! \sqrt{a} - \sin^{-1}\! \sqrt{y}
   -(1\!-\!2a)\sqrt{a(1\!-\!a)}\right] ~ & \quad  \alpha=\frac{3}{2}\,,\\[4mm]
 \tfrac{1}{3}N\left\{2+2\ln [2(1\!-\!y)]-\frac{1}{1-y}
   -a^2(3\!-\!2a)\left[\frac{1-2y}{y(1-y)}
     + 2 \ln \frac{(1-y)}{y}\right]\right\}  ~ & \quad  \alpha=2\,.
\end{cases}
\end{align}
The main qualitative feature of $T_{\rm pol}(y)$ is that it is maximal for
$y\to 0$ and decreases to 0 as $y\to \frac{1}{2}$.  The $N$ dependence of the
maximal polarization time may be found by setting $y=a=\alpha/(2N)$ in
Eq.~\eqref{Tp} and keeping the dominant contribution.  This gives:
\begin{align*}
  T_{\rm pol} \simeq
\begin{cases}
\frac{1}{4}\pi^2N \approx 2.467 N &\quad\alpha=\frac{1}{2}\,,\\[1mm]
2 \ln 2 N\approx 1.386 N &\quad \alpha=1\,,\\[1mm]
N &\quad \alpha=\frac{3}{2}\,,\\[1mm]
\frac{1}{3}(1+2\ln 2)N \approx 0.795 N &\quad \alpha=2;
\end{cases}
\end{align*}
that is, the maximal value of the polarization time scales linearly with $N$.

\begin{figure}[ht]
  \centerline{
    \subfigure[]{\includegraphics[width=0.305\textwidth]{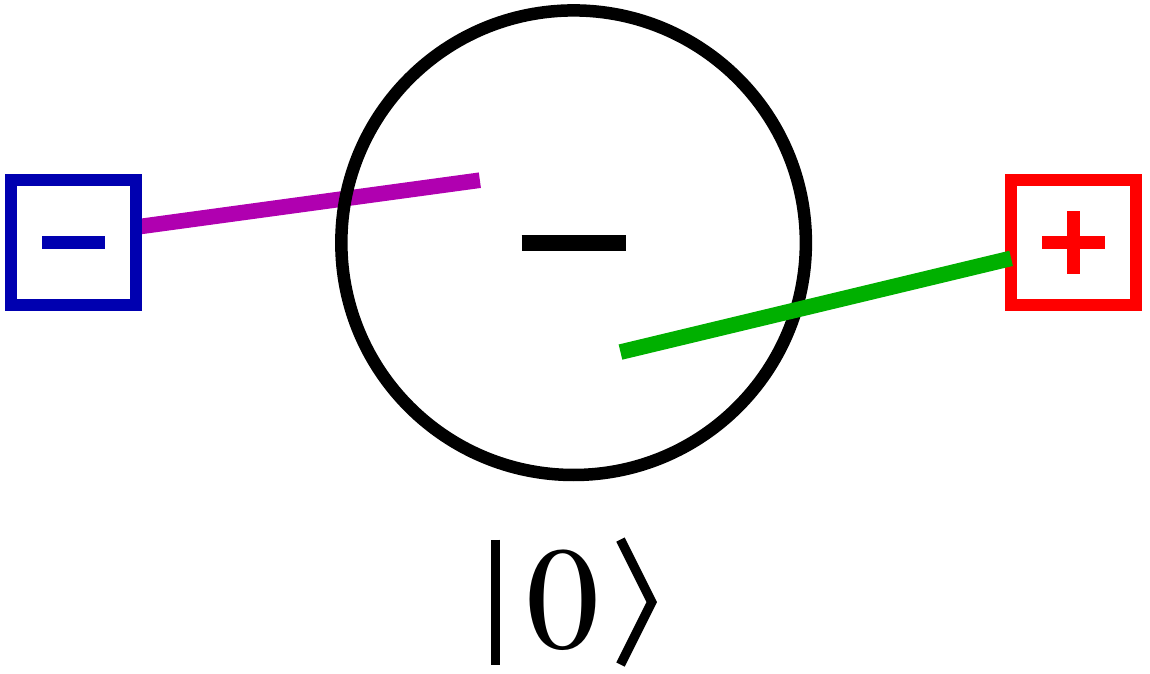}}\qquad
    \subfigure[]{\includegraphics[width=0.305\textwidth]{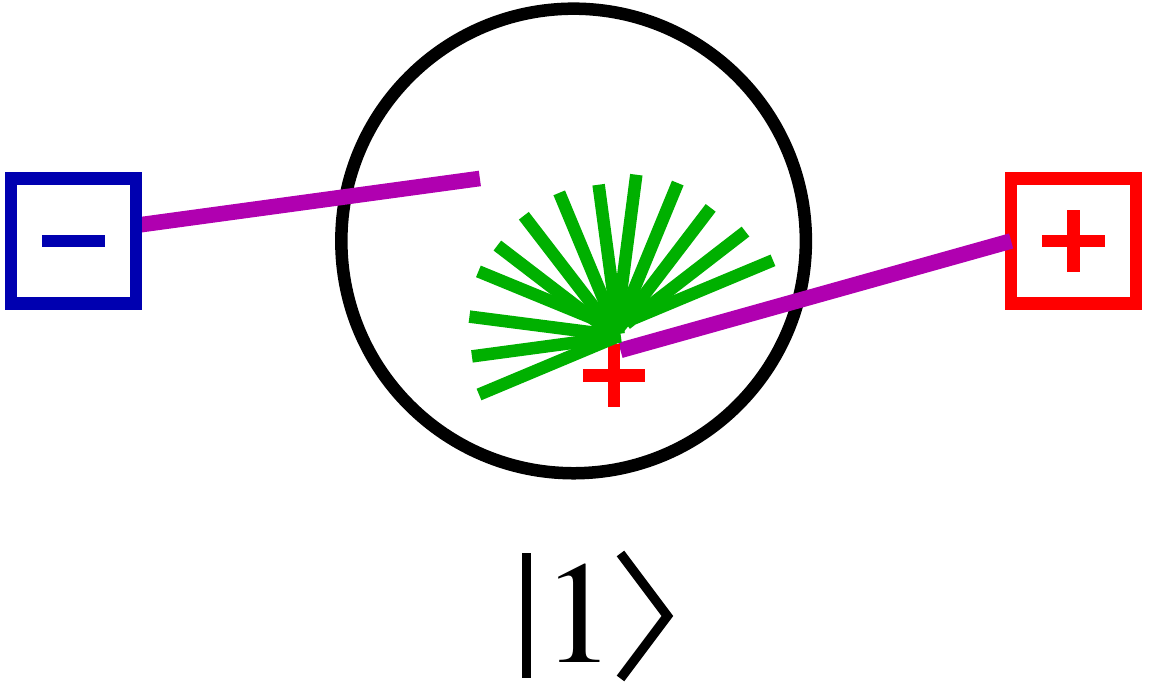}}\qquad
    \subfigure[]{\includegraphics[width=0.305\textwidth]{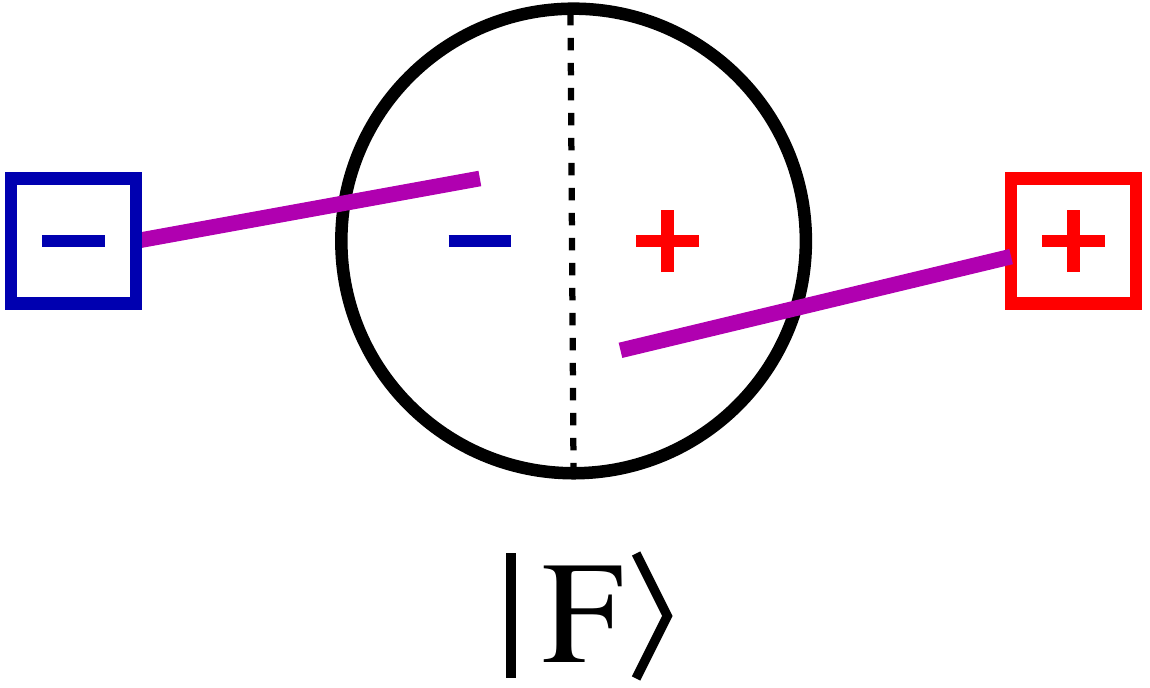}}}
  \caption{\small The three states in the reduced state space of the
    complete-graph system with a single link from each news source: (a) the
    initial consensus state $|0\rangle$.  Active and inactive links are shown
    in green and magenta. (b) The excited state $|1\rangle$, where the single
    voter linked to the $+$ news source changes its opinion to $+$ and now
    disagrees with all its $N-1$ neighbors, leading to many active links.
    (c) The final polarized state $|F\rangle$; here the active links are not
    shown.}
 \label{CG1}  
\end{figure}

However, the polarization time grows faster than linearly in $N$ for
sufficiently small $\alpha$, corresponding to weak news sources.  To
understand the behavior in this limit, it is instructive to consider the
extreme limit where each news source is connected to a single voter
(Fig.~\ref{CG1}).  Suppose that the population starts in the $-$ consensus
state.  At some point, an ``informed'' voter (the one linked to the $+$ news
source) changes its opinion from $-$ to $+$ by interacting with the news
source.  When this happens, this informed voter now disagrees with all its
neighbors.  From this excited state, subsequent opinion changes are primarily
caused by disagreeing voters within the complete graph because links between
voters are numerous and there is only one link to the news source.  Thus the
voters undergo classic VM dynamics, as long as there is any
disagreement.

\begin{figure}[ht]
\centerline{\includegraphics[width=0.375\textwidth]{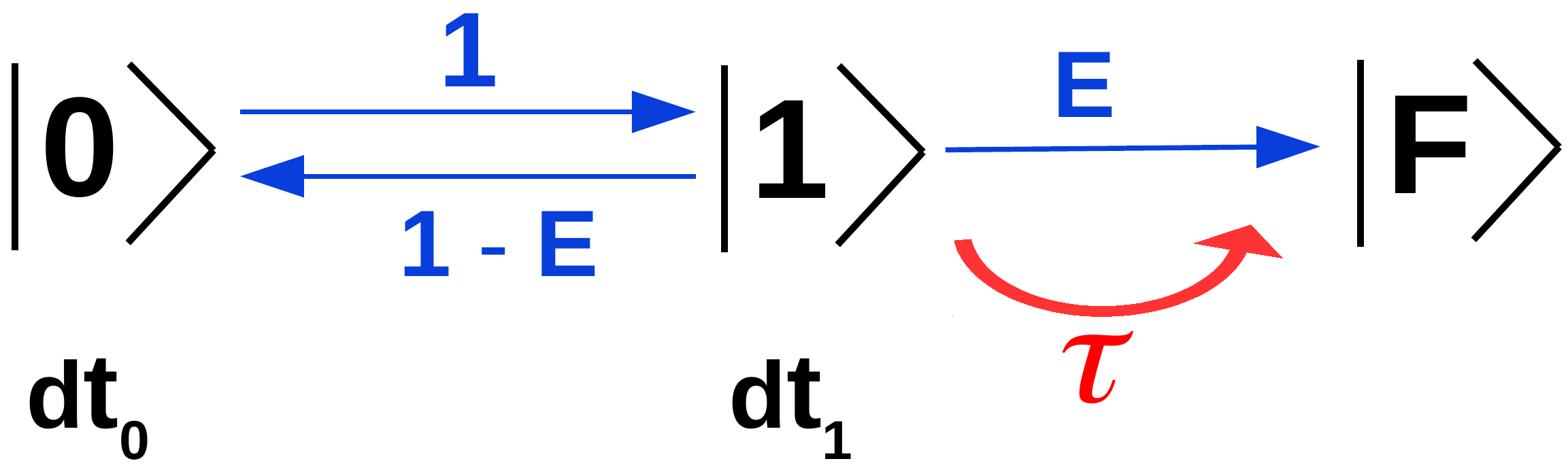}}
\caption{\small The transition rates and transition times in the reduced
  system.}
\label{state}
\end{figure}

Within this picture, we can reduce the dynamics to a three-state space
(Fig.~\ref{state}): the state $|0\rangle$, corresponding to $-$ consensus
($x=0$), in which only the news source influences the voters, the final
polarized state $|F\rangle$ ($x=\frac{1}{2}$), and the excited state
$|1\rangle$, in which one voter in the complete graph has the $+$ opinion.
As indicated in Fig.~\ref{CG1}(b), the news source has a negligible influence
on this excited state.  After this reduction, it is straightforward to
compute the time to reach the polarized state $|F\rangle$ starting from the
initial consensus state $|0\rangle$ by applying first-passage
ideas~\cite{R01}.  Starting from $|0\rangle$, the state $|1\rangle$ is
necessarily reached, so the transition probability from $|0\rangle$ to
$|1\rangle$ equals 1.  Similarly, $E=\frac{2}{N}$ is the probability to reach
polarized state $|F\rangle$ from $|1\rangle$ by VM dynamics (that is, one $+$
voter initially and $\frac{N}{2}$ $+$ voters in the final state).  This
portion of the dynamics coincide with the VM because the news sources play no
role.

We define $T_0$ and $T_1$ as the first-passage times to reach the polarized
state from the initial states $|0\rangle$ and $|1\rangle$ respectively.
These first-passage times satisfy
\begin{align}
\begin{split}
 \label{T010}
 T_0&=dt_0+T_0' \,,\\
  T_1 &= E \tau\ + (1\!-\!E)(dt_1+T_1)\,,
\end{split}
\end{align}
where, from Eq.~\eqref{eqrates},
\begin{align*}
  dt_0&=1/[r^+(0)\!+\!r^-(0)]=2/\alpha\,,\\
        dt_1&=1/\big[r^+(\tfrac{1}{N})\!+\!r^-(\tfrac{1}{N})\big]\approx 1\,,
\end{align*}
are the transition times to leave the states $|0\rangle$ and $|1\rangle$,
respectively, and $\tau=2N(1-\ln 2)$ is the conditional time to reach the
final state $|F\rangle$ from $|1\rangle$ by VM dynamics
(\ref{conditional-polarization-vm}).  Solving Eqs.~\eqref{T010} gives,
\begin{align}
\label{Tp2}
  T_{\rm pol}&\equiv T_0=\tau +\frac{dt_0}{E}+\frac{(1-E) dt_1}{E}
  \approx N\left(\tfrac{5}{2}-2\ln 2\right) + \frac{N }{\alpha}.
\end{align}
Thus the polarization time scales linearly with $N$, unless $\alpha\to 0$.
This limit of $\alpha\to 0$ is achieved, for example, when a single link
connects the news source to voters.  In this case,
$\alpha=p\ell/(1-p)=p/[N(1-p)]$, which gives $T_{\rm pol} \sim N^2(1-p)/p$.

\subsubsection{Asymmetrically connected news sources}

\begin{figure}[ht]
  \centerline{    \includegraphics[width=0.45\textwidth]{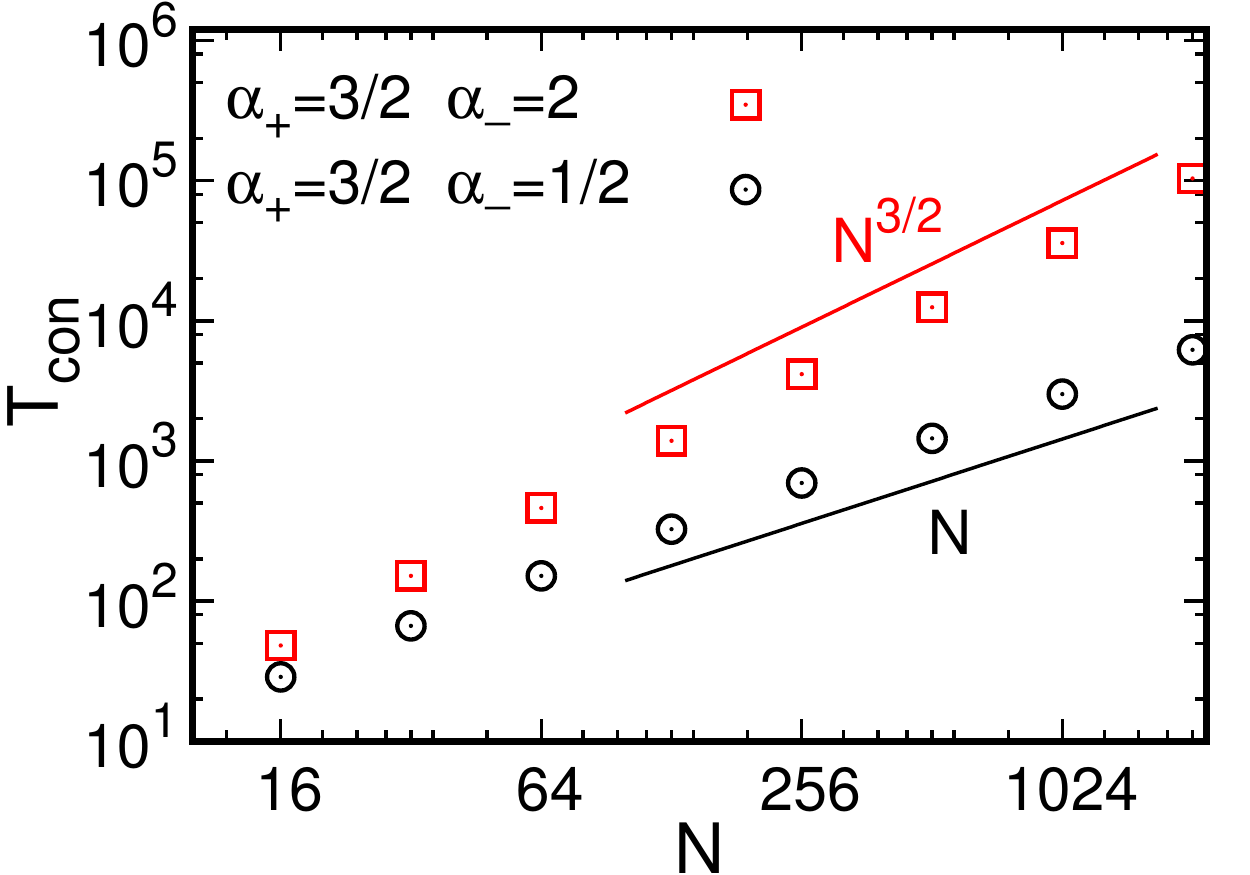}  }
  \caption{\small The consensus time versus $N$ for voters influenced by two
    news sources with unequal link densities.  The data are from simulations
    over $10^4$ realizations.  The lines are guides for the eye. }
\label{Tc-vs-N-unequal}  
\end{figure}

When the number of links from the two news sources differ, the density $x$
now diffuses in an asymmetric logarithmic potential.  To achieve consensus,
$x$ has to surmount one of the potential barriers, either at $a_-$ or at
$1-a_+$, with respective barrier heights $\alpha_+ \ln N$ and
$\alpha_- \ln N$.  Again from Kramers' theory, the dominant contribution to
the consensus time scales exponentially in the lowest barrier height, as long
as the barrier height grows at least as fast as $\ln N$.  Thus the consensus
time scales as $N^\alpha$, with consensus time exponent now given by
$\alpha=\min(\alpha_+,\alpha_-,1)$.  To test this hypothesis, we show the $N$
dependence of $T_{\rm con}$ from simulations in Fig.~\ref{Tc-vs-N-unequal}
for two combinations of unequal link densities.  For
$(\alpha_+,\alpha_-)=(\frac{3}{2},\frac{1}{2})$, the consensus time scales
linearly with $N$, while for $(\alpha_+,\alpha_-)=(\frac{3}{2},2)$, the
consensus time scales as $N^{3/2}$, as we expect.  In the simulations, we fix
$p=\frac{2}{3}$ and $\ell_+=\frac{3}{4}$ to give $\alpha_+=\frac{3}{2}$, and
then use $\ell_-=\frac{1}{4}$ to give $\alpha_-=\frac{1}{2}$ and $\ell_-=1$
to give $\alpha_-=2$.

Another basic characteristic of the collective opinion state is its
distribution.  The opposing nature of the two news sources drives the
population to a steady state in the long-time limit.  We obtain the
steady-state opinion distribution,
$P_{\rm ss}(x)\equiv P(x,t\rightarrow \infty)$, by setting
${\partial P}/{\partial t}=0 $ in the Fokker-Planck equation \eqref{FP} and
then solving.  To have a well-posed problem, we need to specify the boundary
conditions.  The appropriate conditions are reflection at $x=a_-$ and
$x=1-a_+$ because for all $\alpha>0$, the endpoints are not fixed points of
the stochastic dynamics.  Solving this Fokker-Planck equation and imposing
normalization, $\int^{1-a_+}_{a_-}P_{\rm ss}(x)dx=1$, we obtain
\begin{align}
  P_{\rm ss}(x)=\frac{x^{\alpha_+-1}(1-x)^{\alpha_--1}}
 {{B}\left[1-a_+;\alpha_+,\alpha_-\right]-{B}\left[a_-;\alpha_+,\alpha_-\right]}\,,
 \label{ss}
\end{align}
where $B(x;y,z)$ is the incomplete beta function~\cite{dilog}.

\begin{figure}
  \centerline{\includegraphics[width=0.45\textwidth]{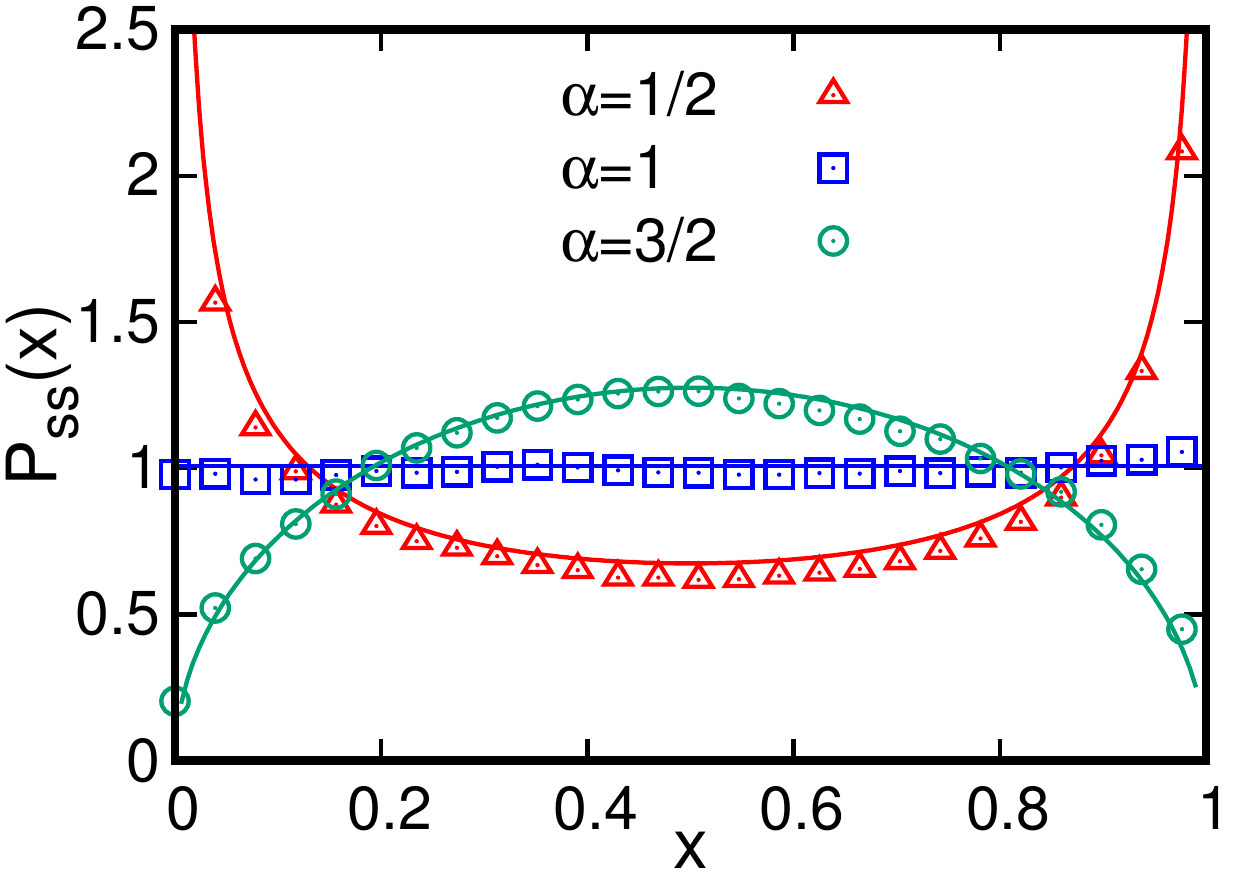}}
  \caption{\small Steady-state distribution of $x$ for 128 voters.  The
    curves are the predictions of Eq.~\eqref{ss} and symbols represent
    simulations.}
\label{distribution}  
\end{figure}

In the symmetric case, $\ell_{\pm}=\ell$, this distribution reduces to
$P_{\rm ss}(x) \propto [x(1-x)]^{\alpha-1}$, which undergoes a bimodal to
unimodal transition as $\alpha$ passes through 1 (Fig.~\ref{distribution}).
In this figure, we fix $\ell=1$ and use $\alpha =p\ell/(1-p)$ with
appropriate values of $p$ to give $\alpha=\frac{1}{2}, 1$, and $\frac{3}{2}$,
and then evolve the system until the steady state is reached (typically for
times greater than $10^6$).  When $\alpha<1$, the distribution has maxima at
$x=0,1$.  That is, for weak news sources, the population typically remains
close to one of the two consensus states.  Conversely, for influential news
sources ($\alpha>1$) the steady-state distribution has a maximum at $x=1/2$,
corresponding to the politically polarized state.  In the marginal case of
$\alpha=1$, all possible opinion states are equally likely.

\section{Voters on a two-clique graph}
\label{twocliquegraph}

\begin{figure}[h]
  \centerline{\includegraphics[width=0.6\textwidth]{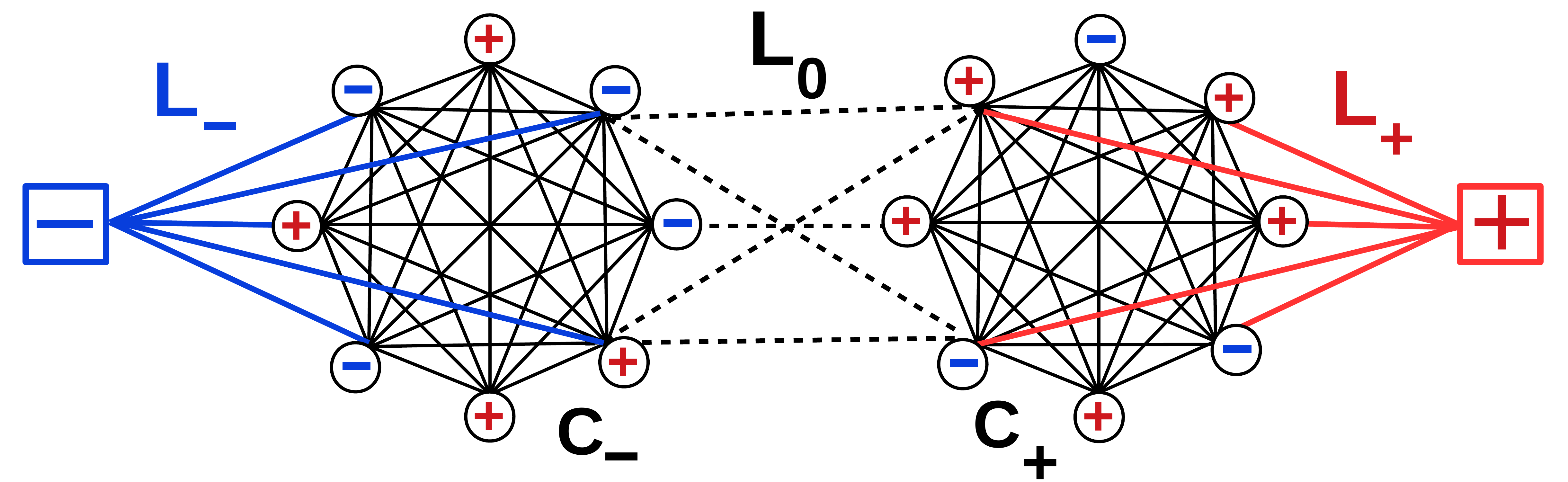}}
  \caption{ The two-clique system, with two opposing news sources (squares)
    and voters (circles).  Each clique contains $N$ voters, with $L_0$ links
    between cliques $C_+$ and $C_-$.}
\label{two-clique}
\end{figure}

We now investigate the influence of two opposing news sources when the voters
reside on a two-clique graph, with $N$ voters in each clique
(Fig.~\ref{two-clique}).  The $+$ news source connects to random voters on
$C_+$ via $L_+$ links and the $-$ news source connects to random voters on
$C_-$ via $L_-$ links.  We write $\ell_{\pm}=L_{\pm}/N$ as the corresponding
link densities.  To simplify matters, we restrict ourselves to the case of
equally connected news sources, $\ell_+=\ell_-\equiv\ell$.  However, the
voter model on the two-clique graph with unequal-size cliques was very
recently investigated in~\cite{GI19}, where a non-monotonic dependence of the
consensus time on interclique density was found.  In our symmetric two-clique
graph, the voters on different cliques are connected by $L_0=N^{\beta}$
interclique links, with $0\leq \beta \leq 2$.  For $L_0\to N^2$, the two
cliques together form a complete graph of $2N$ voters.  We focus on the
interesting (and realistic) case where the cliques are sparsely
interconnected ($\beta\to 0$).

Let $x_1$ and $x_2$ denote the fraction of $+$ opinion voters on clique $C_+$
and $C_-$, respectively, at time $t$.  We represent the state of the system
by the clique densities $(x_1,x_2)$.  Let $r_i^{\pm}(x_1,x_2)$ be the rates
for $x_i$ to change by $\pm \delta x$.  Within the annealed-link
approximation, these rates are (see \ref{2C} for details):
\begin{align}
\label{rates-tc}
r_1^{+}(x_1,x_2)&=\tfrac{1}{2}A\left[Nx_1(1\!-\!x_1)+\ell_0(1\!-\!x_1)x_2\right]+B(1\!-\!x_1) \nonumber\\
r_1^{-}(x_1,x_2)&=\tfrac{1}{2}A\left[Nx_1(1\!-\!x_1)+\ell_0\,x_1(1\!-\!x_2)\right] \nonumber\\
r_2^{+}(x_1,x_2)&=\tfrac{1}{2}A\left[Nx_2(1\!-\!x_2)+\ell_0\,x_1(1\!-\!x_2)\right] \nonumber\\
r_2^{-}(x_1,x_2)&=\tfrac{1}{2}A\left[Nx_2(1\!-\!x_2)+\ell_0(1\!-\!x_1)x_2\right]+Bx_2
\end{align}
where $\ell_0=L_0/N=N^{\beta-1}$ is the number of voters in $C_+$ that link
to a voter in $C_-$ or vice versa, and the coefficients $A$ and $B$ are
\begin{subequations}
\label{coefficients-tc}
\begin{align}
\begin{split}
 A&=\frac{\ell(1-p)N}{(1-p)(N+\ell_0-1)+p} + \frac{(1-\ell)N}{N+\ell_0-1}~,\\[2mm]
 B&=\frac{\ell pN}{2[(1-p)(N+\ell_0-1)+p]}~. 
\end{split}
\end{align}
Ignoring terms of order $1/N$, these coefficients reduce to
\begin{align}
 A\approx \frac{N}{N+\ell_0}\,, \qquad B \approx  \frac{\ell p}{2(1-p)}\,.
\end{align}
\end{subequations}

Let $P(x_1,x_2)\,\delta x^2$ be the probability for the opinion state of the
population to be within a range $\delta x^2$ about $(x_1,x_2)$.  Expanding
the underlying master equation in a Taylor series to second order gives the
two-variable Fokker-Planck equation
\begin{align}
 \frac{\partial}{\partial t} P(x_1,x_2,t)= \sum^{2}_{i=1}\left\{-\frac{\partial }{\partial x_i}[V_iP] + \frac{\partial^2 }{\partial x_i^2}[D_iP]\right\}\,,
 \label{FP2}
\end{align}
where
\begin{align*}
  V_i(x_1,x_2)&=[r_i^+(x_1,x_2)-r_i^-(x_1,x_2)] \delta x\,,\\[1mm]
  D_i(x_1, x_2)&=[r_i^+(x_1,x_2)+r_i^-(x_1,x_2)] (\delta x^2/2)\,.
\end{align*}
The coupling between $x_1$ and $x_2$ arises because a change in $x_1$ alters
$V_2$ and $D_2$, and vice versa for $x_2$.  Because of this complication, an
analytical approach of the full dynamics appears to be challenging.
However, Ref.~\cite{GI19} has made progress in this direction.

\begin{figure}[ht]
  \centerline{\subfigure[]{\includegraphics[width=0.45\textwidth]{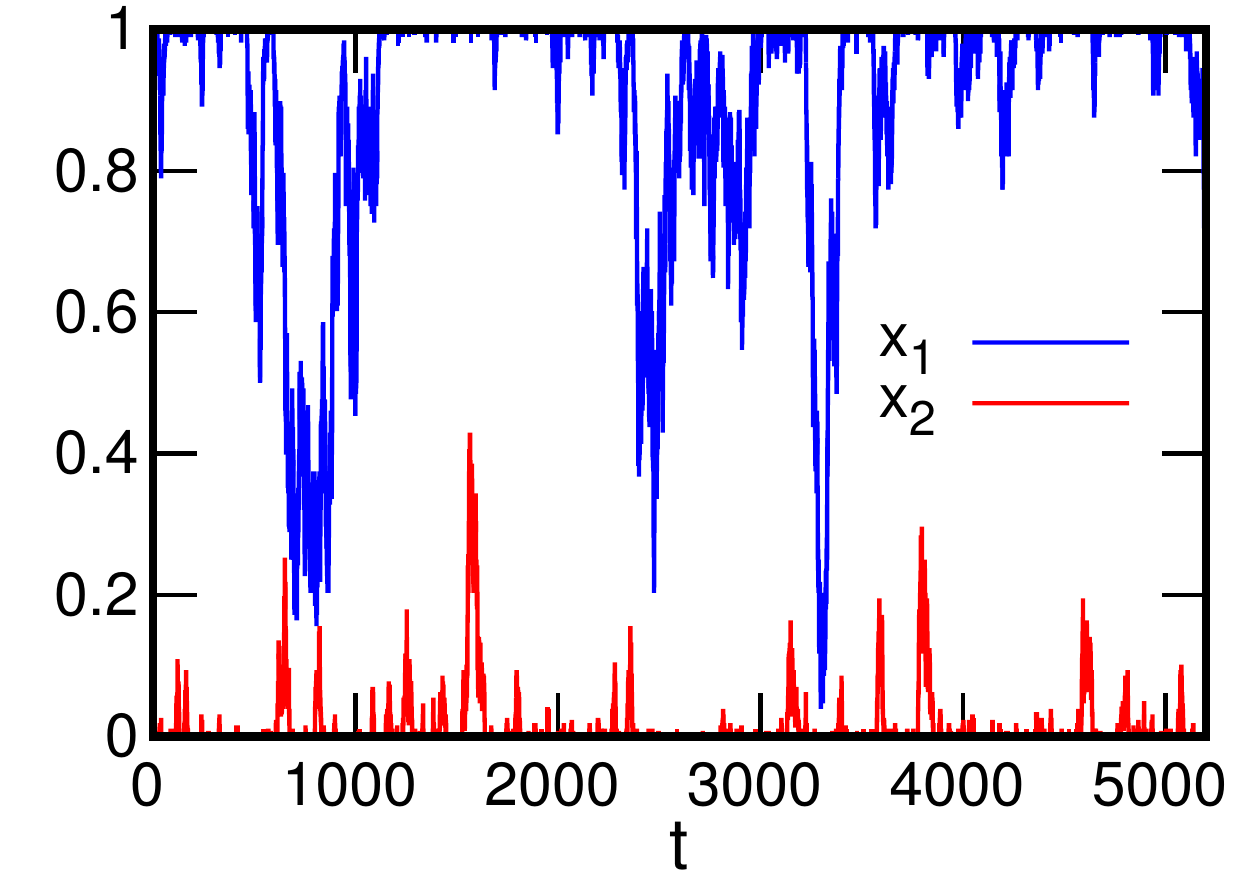}}
  \qquad
    \subfigure[]{\includegraphics[width=0.45\textwidth]{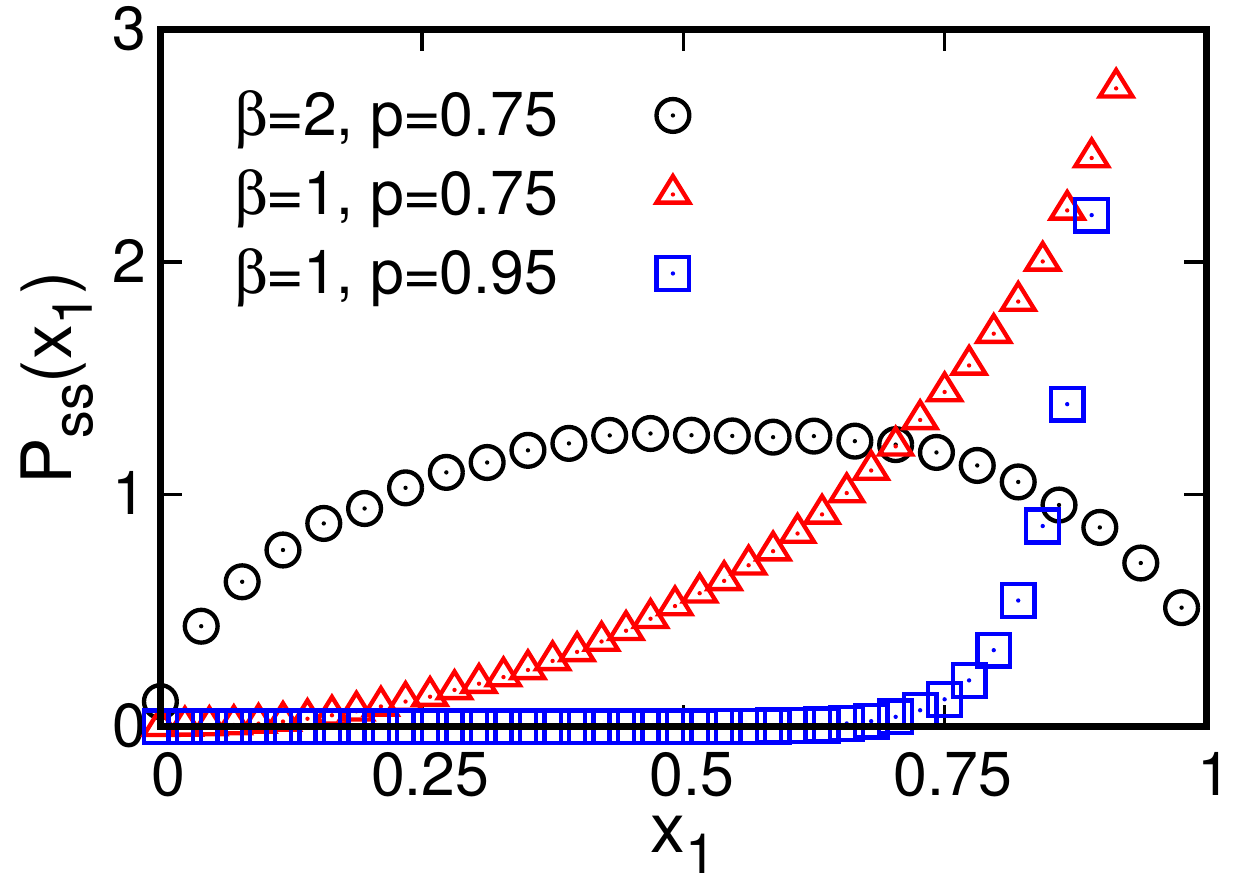}}}
  \caption{(a) Typical trajectories for $x_1$ and $x_2$ in the two-clique
    graph with $N=128$, $\beta=0.5$, and $\alpha=2$ ($\ell=1$ and $p=2/3$).
    (b) Distribution of fraction $x_1$ of $+$ opinion voters on clique $C_+$
    of 128 voters on the two-clique graph, with $\ell=1$.}
\label{two-clique-traj}
\end{figure}

To make progress, it is helpful to first study the time evolution of the
trajectories of $x_1$ and $x_2$ for sparsely connected cliques
(Fig.~\ref{two-clique-traj}(a)).  The population spends a large fraction of
the time in the neighborhood of the state $(x_1\!=\!1, x_2\!=\!0)$, which we
term the \emph{maximally polarized} (MP) state.  The population tends to
remain close to the MP state because: (i) the news sources tend to drive the
clique opinions to this state, and (ii) the transition time to leave the MP
state,
\begin{align*}
 dt_0=\left[r_1^+(1,0)+r_1^-(1,0)+r_2^+(1,0)+r_2^-(1,0)\right]^{-1}\,,
\end{align*}
scales as $N^{1-\beta}$, which becomes large as $\beta\to 0$.  These two
properties is reflected in the steady-state opinion distribution in each
clique (Fig.~\ref{two-clique-traj}(b)).  As shown in the figure, this
distribution becomes more concentrated near $x_1=1$ as either the number of
interclique links is reduced or the interactions with news sources become
stronger.  (By symmetry, the opinion distribution on $C_-$ is concentrated
near $x_2=0$.)~

\begin{figure}[ht]
\centerline{
    \includegraphics[width=0.3\textwidth]{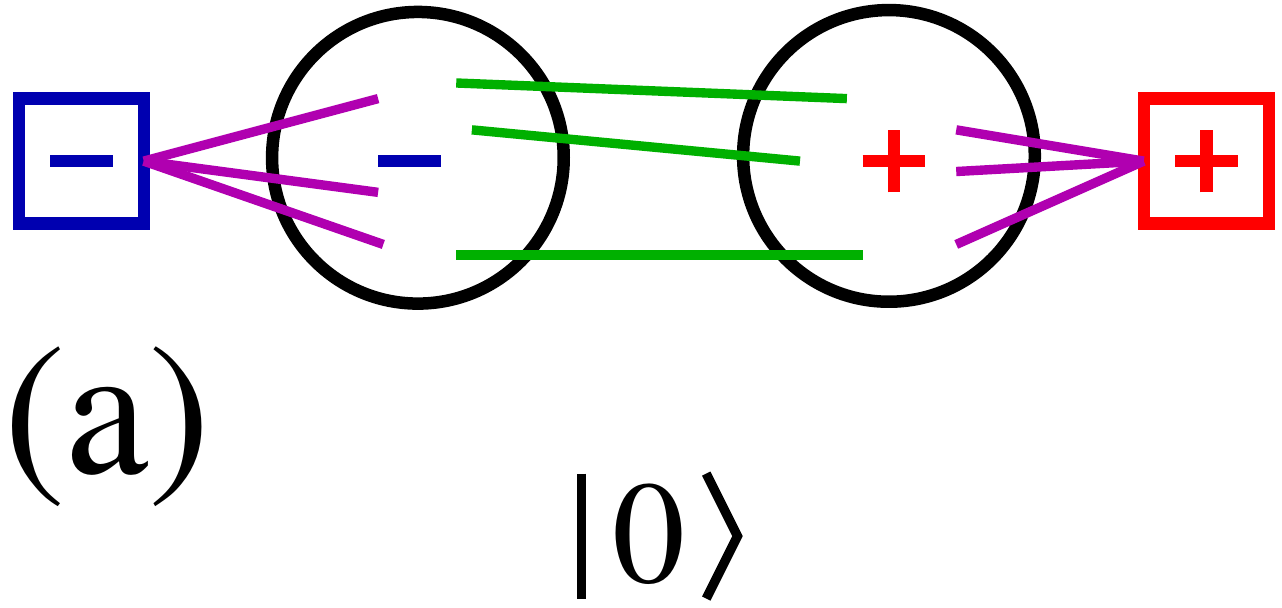}\qquad
    \includegraphics[width=0.3\textwidth]{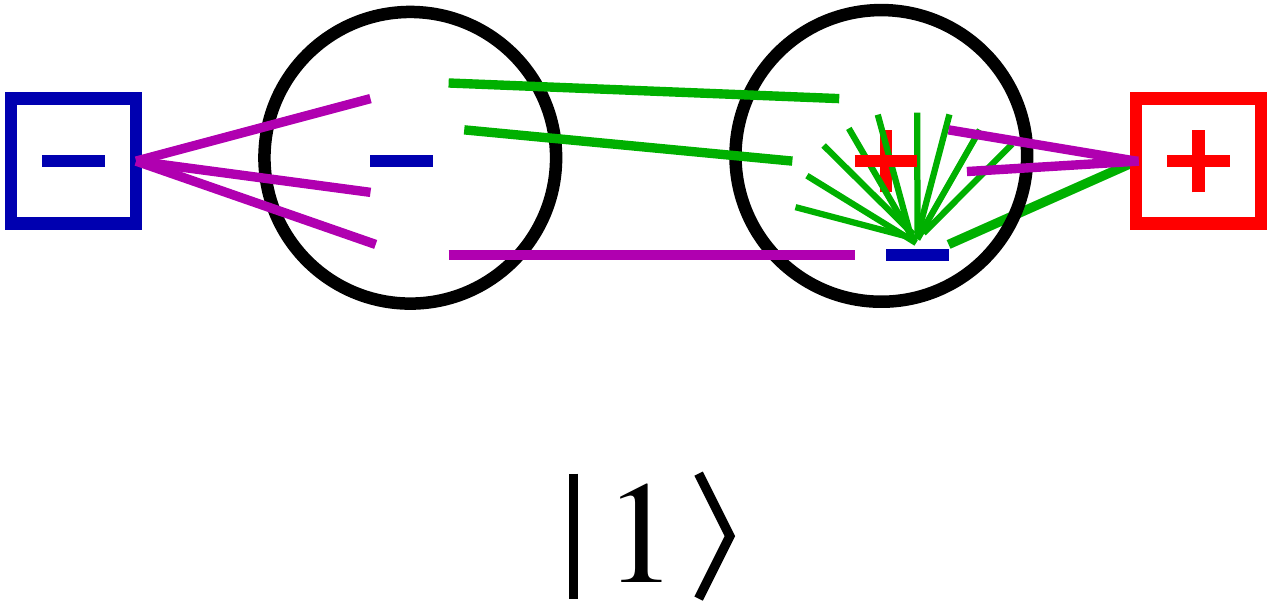}\qquad
    \includegraphics[width=0.3\textwidth]{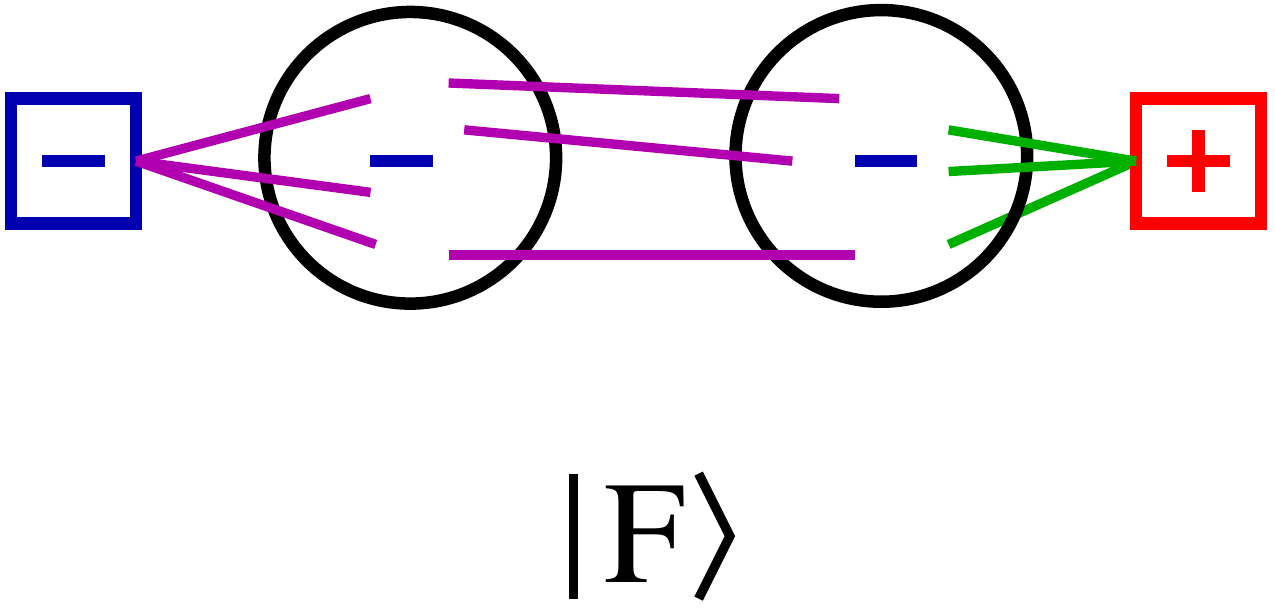}}\bigskip
\centerline{
    \includegraphics[width=0.3\textwidth]{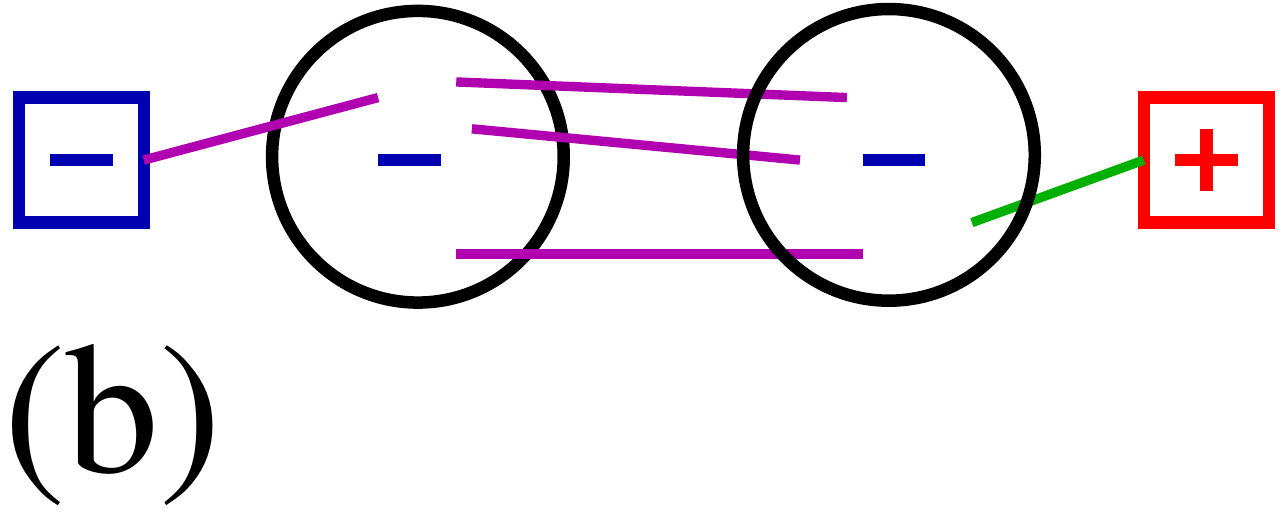}\qquad
    \includegraphics[width=0.3\textwidth]{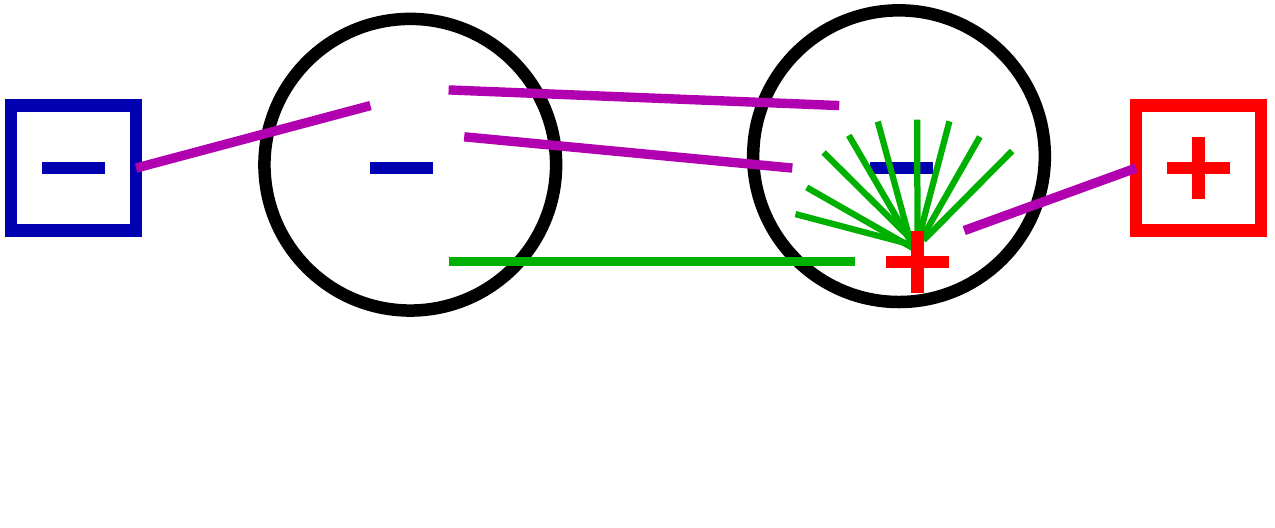}\qquad
    \includegraphics[width=0.3\textwidth]{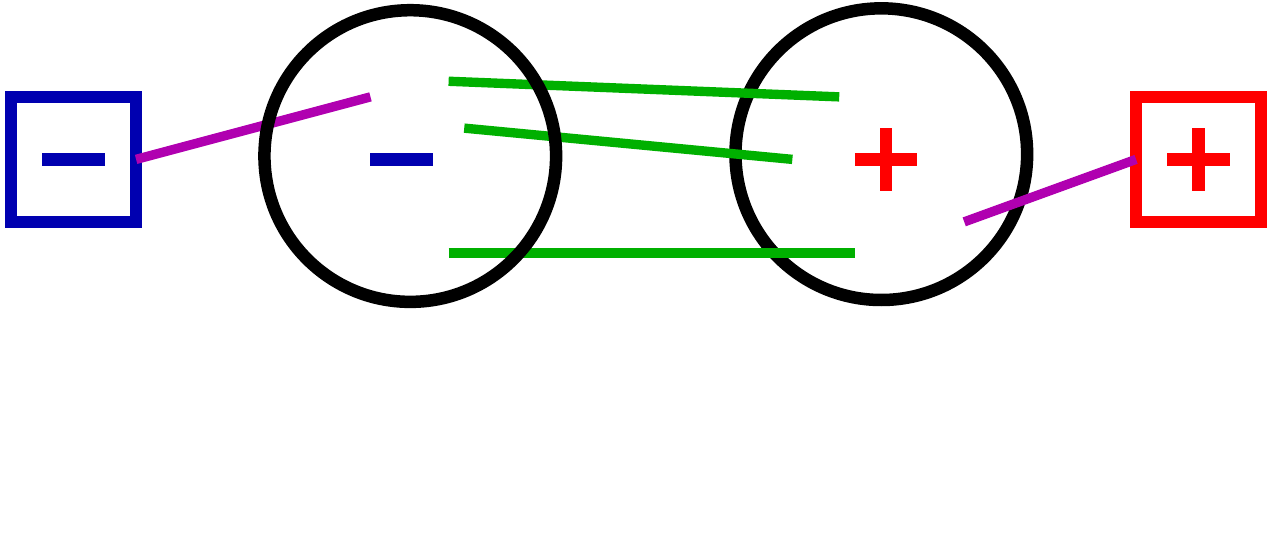}}
  \caption{(a) The reduced states to determine the consensus time for the
    two-clique graph: (left) the initial maximally polarized state
    $|0\rangle$.  (middle) The excited state $|1\rangle$, in which one voter
    in clique $C+$ has changed opinion from $+$ to $-$.  (right) The final
    consensus state $|F\rangle$.  (b) Reduced states of the two-clique graph
    to determine the polarization time: (left) the initial consensus state
    $|0\rangle$.  (middle) The excited state $|1\rangle$, in which one voter
    in clique $C+$ has changed opinion from $-$ to $+$.  (right) The final
    maximally polarized state $|F\rangle$.  }
\label{2Cr}
\end{figure}

In the limit of sparsely connected cliques, we may again reduce the state
space (Fig.~\ref{2Cr}(a)), analogous to the construction given in
Figs.~\ref{CG1} and \ref{state}, to determine the consensus time for the
complete-graph system.  First, note that the $N^\beta$ interclique links are
relevant only in the MP state.  In all other opinion states, the dynamics is
controlled by the $\frac{1}{2}N(N-1)\gg N^\beta$ intraclique links.  Thus we
can view the system as being comprised of two isolated cliques, with one
clique inactive (where all voters agree with the news source) and the other
clique, where the voters are not in consensus, active.

Starting in the MP state, suppose that one voter in $C_+$ changes opinion
from $+$ to $-$ due to an interaction with a $-$ voter in $C_-$.  The
population is now in the excited state $\big(1-\frac{1}{N},0\big)$ where the
$-$ voter in $C_+$ differs with the rest of its $N-1$ neighbors.  Because
interclique links are sparse $(\beta \to 0)$ and the number of intraclique
links between voters with differing opinions in $C_+$ is of order $N$, the
opinion dynamics is driven by the latter class of links.  For the active
clique $C_+$, there are two possible outcomes starting from the excited state
$\big(1-\frac{1}{N},0\big)$.  Either $C_+$ returns to $+$ consensus (and the
full system returns to the MP state) or $C_+$ reaches $-$ consensus.  We
again visualize the MP state $(1,0)$ as $|0\rangle$, the excited state
$\big(1-\frac{1}{N},0\big)$ as $|1\rangle$, and the $-$ consensus state
$(0,0)$ as the final state $|F\rangle$ (Fig.~\ref{2Cr}(a)).  With these
reduced states, we obtain the consensus time by the same calculation as that
given in the previous subsection to determine the polarization time for the
complete graph.

Starting from the state $|0\rangle$, the system moves to state $|1\rangle$
after a transition time $dt_0=N^{1-\beta}$.  The time to reach the
final consensus state starting from $|0\rangle$ satisfies
\begin{subequations}
\begin{align}
\label{tc-con-1}
  T_0=dt_0 +  T_1\, .
\end{align}
where $T_1$ is the time to reach consensus from the initial state
$|1\rangle$.  Substituting the state $|1\rangle$ densities
$(x_1,x_2)=\big(1-\frac{1}{N},0\big)$ into the rates \eqref{rates-tc}, the
transition time to leave state $|1\rangle$ is
$dt_1\approx (1+\ell_0)^{-1}$.  Starting from $|1\rangle$, the
probability for the active clique to reach the state $|2\rangle$ is
$E=1-E_+\big(1\!-\!\frac{1}{N}\big)$, with $E_+(y)$ given in
Eq.~\eqref{exit-amass}.  We also define the mean conditional time to reach
$|2\rangle$ from $|1\rangle$ as $\tau$.  Then $T_1$ satisfies
\begin{align}
\label{tc-con-2}
T_1=   E \tau + (1-E)(dt_1+T_0)\, .
\end{align}
\end{subequations}
Solving these equations gives
\begin{align}
  \label{tc-twoclique}
 T_0\equiv T_{\rm con}=\tau+\frac{dt_0}{E}+\frac{(1-E) dt_1}{E}\, .
\end{align}
In the limit $\beta \to 0$, $\tau$ is subdominant (see
\ref{conditional-consensus}) and $dt_1\ll dt_0$, so that the third term is
also subdominant.  Thus keeping only the dominant term $dt_0/E$, the
consensus time has scaling behavior:

\begin{align}
\label{Tc-N-tc}
T_{\rm con}\sim
\begin{cases}
   N^{2-\beta} \quad &   \alpha< 1\, ,\\
   N^{2-\beta} \ln N   \quad &  \alpha=1\, ,\\
   N^{1+\alpha-\beta}  \quad &  \alpha> 1\, .
\end{cases}
\end{align}

\begin{figure}[ht]
  \centerline{
    \subfigure[]{\raisebox{1.5mm}{\includegraphics[width=0.4\textwidth]{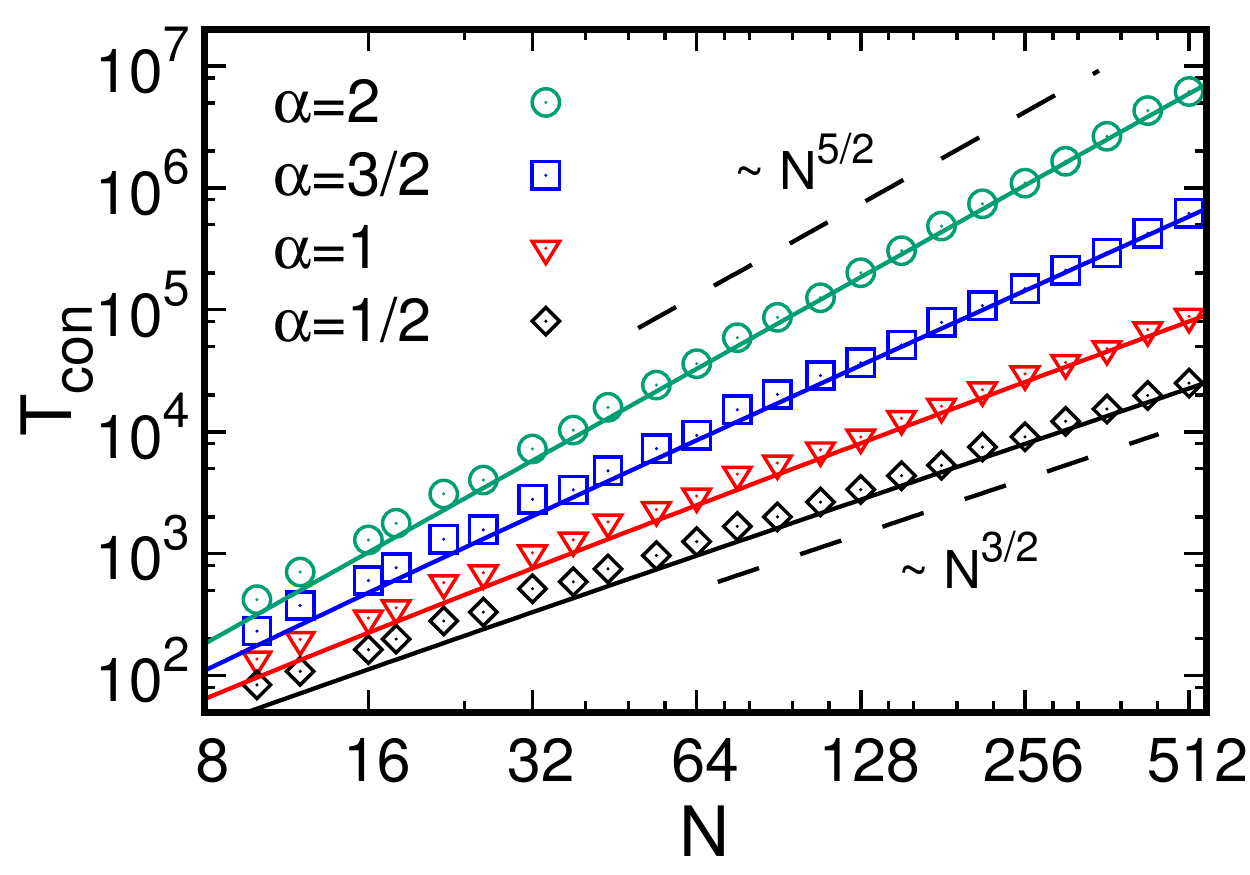}}}\qquad
    \subfigure[]{\includegraphics[width=0.41\textwidth]{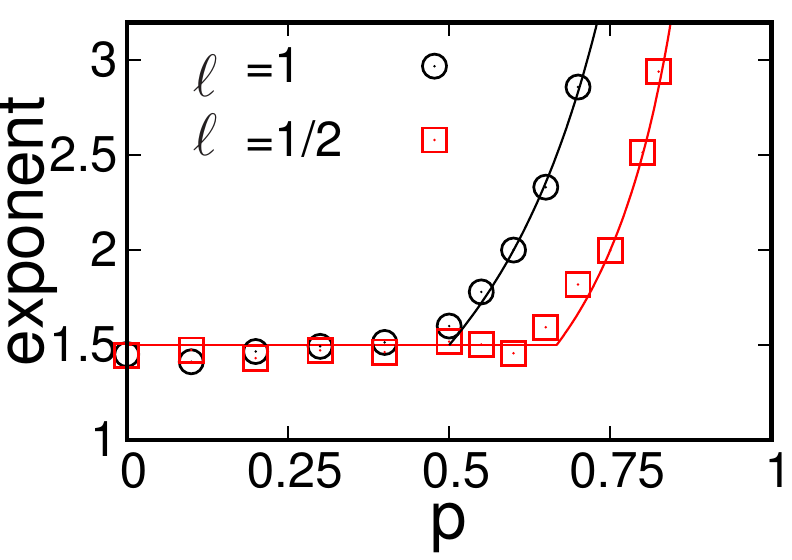}}}
  \caption{\small (a) Consensus time versus $N$ for voters on a two-clique
    graph.  The solid lines give $dt_0/E$ from Eq.~\eqref{tc-twoclique} and
    the symbols represent simulations over $10^3$ realizations.  (b)
    Consensus time exponent as a function of $p$.  The symbols are the
    simulation results and the curves are the predictions of
    Eq.~\eqref{Tc-N-tc}.}
\label{consensus-twoclique}  
\end{figure}

The consensus time exponent increases as interclique links become more rare
and also as the influence of news sources increases beyond $\alpha\geq 1$
(Fig.~\ref{consensus-twoclique}(a)).  The data in this figure corresponds to
fixed $\beta=\frac{1}{2}$ and $\ell=1$, and $p$ is varied to give the
$\alpha$ values shown.  Figure~\ref{consensus-twoclique}(b) shows the
consensus time exponent as a function of $p$ for fixed $\ell$, as well as the
comparison with our basic prediction Eq.~\eqref{Tc-N-tc}.  Our results for
the consensus time are consistent with previous studies~\cite{GI19,Masuda3}
of the VM on the two-clique graph.
 
For the polarization time, we again use a reduced state-space approach,
analogous to that developed to derive Eq.~\eqref{Tp2} for the complete-graph
system.  For concreteness and simplicity, we start the system in the $-$
consensus state $(0,0)$ (Fig.~\ref{2Cr}(b)) and determine the time to reach
the $(1,0)$ MP state in the interesting limit of weak news sources and weak
intraclique connections.  We now denote the $(0,0)$ consensus state as
$|0\rangle$, the excited state $\big(\frac{1}{N},0\big)$, where a single
voter in clique $C_+$ has changed opinion, as $|1\rangle$, and the MP state
as $|F\rangle$, respectively (Fig.~\ref{state}).  In state $|0\rangle$, the
opinion change that leads to state $|1\rangle$ is caused only by the link
between a voter and the $+$ news source.  In $|1\rangle$, subsequent opinion
changes occur by classic VM dynamics because interclique links significantly
outnumber all other links and the effect of the latter can be ignored.

Referring to Fig.~\ref{2Cr}(b), let $dt_0$ and $dt_1$ denote the
transition times out of the states $|0\rangle$ and $|1\rangle$ respectively.
These transition times are the inverse of the sum of the rates out of these
states.  To obtain $dt_0$ we substitute $(x_1,x_2)=(0,0)$ into the
rates \eqref{rates-tc} and obtain, after straightforward steps,
$dt_0=2(1-p)/p\ell$.  Similarly for $dt_1$, we substitute
$(x_1,x_2)=\big(\frac{1}{N},0\big)$ into \eqref{rates-tc} and ultimately
obtain $dt_1\approx 1$ for large $N$ and $\ell\to 0$.  Let $E$ denote
the exit probability to reach $|F\rangle$ starting from $|1\rangle$ without
reaching $|0\rangle$; thus $1-E$ is the probability to reach $|0\rangle$
without reaching $|F\rangle$.  Because the opinions evolve according to the
classic VM dynamics when the system is in state $|1\rangle$,
$E=1-\frac{1}{N}$ and the conditional time to reach $|2\rangle$ from
$|1\rangle$ is $\tau=2N$, (\ref{conditional-consensus}).  Let $T_0$ and $T_1$
be the first-passage times to $|F\rangle$ starting from the initial states
$|0\rangle$ and $|1\rangle$.  These first-passage times again satisfy
Eqs.~\eqref{T010} whose solution now is
\begin{align}
  T_{\rm pol}=\tau+\frac{dt_0}{E}+\frac{(1-E) dt_1}{E}
  \approx 3N  + \frac{2N}{\alpha}~.
 \label{Tp2-tc}
\end{align}
The main message from this result is that as soon as the news sources connect
to a non-vanishing fraction of the population, the polarization time is of
order $N$, and is generally much smaller that the consensus time when
$\alpha$ is large.

\section{Summary}
\label{conclusions}

We introduced an opinion dynamics model where individuals that change their
opinions by VM dynamics are also influenced by two fixed and
opposing news sources.  Our model is motivated by the current polarized
political state in Europe and the US~\cite{AG05,BG08,FA08,S10,P13,IW14,pew},
as well as by the recent emergence of biased news sources that promulgate
fixed political viewpoints~\cite{IH09,L13,MY17}.  Our interest was to
investigate the consequences of political polarization, which seem to be
largely influenced by these types of news sources.  In the VM
framework, the two news sources are zealots that never change their opinion,
but which influence the opinions of individual voters.  Voters, on the other
hand, may consult either news sources or neighboring voters to update their
opinion state.  The strength of the news sources is quantified by a single
parameter $\alpha$, which encapsulates their degree of connection to the
population and the relative likelihood that a voter consults a news source
rather than a fellow voter.  We developed a general framework to understand
the rich dynamics of this model.

Our modeling relies on using highly idealized social networks.  The two
examples that we studied were the complete graph of $N$ voters and, more
realistically, the two-clique graph with $N$ voters in each clique.  The
primary reason for this extreme level of idealization is to formulate
analytically tractable models.  In the complete graph, the news sources
connect either to disjoint or to random voters; both subcases gave virtually
identical results.  In the two-clique graph, each news source connects to
voters in disjoint cliques.  With this modeling perspective, we can
understand many properties of the opinion dynamics analytically.  The
analytical approach also allows us to develop insights that would be
extremely difficult to reach through numerical simulations of our model on
realistic social networks.

We studied basic characteristics of the collective opinion state including:
the exit probability, the consensus time, and the polarization time.
Generally, the consensus time increases while the polarization time decreases
as the news sources becomes more influential.  This behavior can be
understood in terms of a diffusion-like picture for the opinion evolution.
For the complete graph, the fraction $x$ of $+$ voters can be viewed as an
effective particle that undergoes convection-diffusion in the interval
$[0,1]$ in the presence of an effective potential.  Reaching consensus
corresponds to the effective particle surmounting the potential barriers near
$x=0$ or $x=1$, while reaching the polarized state corresponds to the
effective particle being pushed to the minimum of the potential.  This
potential picture explains why the consensus time is much longer than the
polarization time.  This disparity was also reflected in the steady-state
opinion distribution, which undergoes a transition from a bimodal to a
unimodal state as the influences of news sources is increased.

The existence of an effective potential implies that the magnetization,
namely, the difference in the fraction of $+$ and $-$ voters, is not
conserved by the dynamics.  In previous studies of variants of the VM with
non-conserved magnetization, the consensus time was found to grow faster than
any power law in $N$ (see, e.g., \cite{LR07,LSB09,Bhat}).  In constrast, in
this work the effective potential at the boundaries scales logarithmically in
$N$, which leads to a power-law dependence of the consensus time on $N$, with
a non-universal exponent.

We also found that voters on a two-clique graph are driven to a maximally
polarized state in which voters on two different cliques independently reach
unanimity but in opposite opinion states.  The driving mechanism towards this
state becomes stronger either when the number of interclique links is reduced
or when the influence of news sources is increased.  Weakly interconnected
societies are very common around us because of social segregation,
geopolitics (e.g., countries, states, etc.), and cultural differences
(language, religion, etc.).  All of these factors contribute to political
polarization, and our modeling seems to capture an essence of this
phenomenology.  It would be worthwhile to allow the network itself to evolve
to mimic the feature that societies are currently tending to increased
fractionation.
 
We thank Mirta Galesic for helpful discussions and also gratefully
acknowledge NSF financial support from grant DMR-1608211.
 
\bigskip
\appendix

\section{Transition rates}

To determine the transition rates $r^{\pm}(x)$, we need: (a) the elemental
time step $\delta t$ for a single update, and (b) the probabilities
$q^{\pm}(x)$ for $x$ to change by $\pm \delta x$ in an update.  In terms of
these quantities, the rates $r^{\pm}(x)$ are
\begin{align}
 r^+(x)=\frac{q^+(x)}{\delta t}\, , \qquad r^-(x)=\frac{q^-(x)}{\delta t}\,.
 \label{rate1}
\end{align}
The probability for $x$ to not change in an update is
$1-[r^+(x)+r^-(x)]\delta t=1-q^+(x)-q^{-}(x)$.

\subsection{Voter on a complete graph}
\label{CG}

We first determine the rates for the VM on a complete graph of $N$ voters,
where the elemental time step is $\delta t=2/N$.  To find the probabilities,
$q^{\pm}(x)$, first consider $q^{+}(x)$.  For $x$ to increase, a $-$ voter
has to be selected that then adopts the opinion of a neighboring $+$ voter.
The probability to pick a $-$ voter is $N^-/N=(N-N^+)/N$, where $N^+=N x $ is
the number of $+$ voters. The selected $-$ voter has $N-1$ neighbors, of
which $N^+$ have opinion $+$.  The probability for the $-$ voter to pick a
$+$ neighbor is $N^+/(N-1)$.  Therefore the probability for $x$ to increase
by $\delta x$ is,
\begin{align}
 q^+(x)=\frac{N-N^+}{N} \frac{N^+}{N-1} =\frac{x(1-x)}{1-1/N}= q^-(x)\,.
 \label{q-vm}
\end{align}
The last equality follows from the $+ -$ symmetry of the VM. 
 
Substituting $q^{\pm}(x)$ and $\delta t=2/N$ into Eq.~\eqref{rate1}, the
rates $ r^{\pm}(x)$ are
\begin{align}
 r^{\pm}(x)=\frac{Nx(1-x)}{2(1-1/N)}\approx \frac{Nx(1-x)}{2}\,,
 \label{rates-votemrmodel}
\end{align}
as quoted in Eq.~\eqref{rates-vm}.

\subsection{Voters on a complete graph influenced by two news sources}
\label{CG2}

Now consider voters on a complete graph that are additionally influenced by
two news sources (Sec.~\ref{completegraph}). The system consists of $N+2$
agents: $N$ voters and two news sources, so that the elemental time step is
$\delta t= 2/(N+2)$.  We now determine the probability $q^+(x)$ for a voter
to adopt the opinion of a neighboring $+$ voter.  The probability to pick a
$-$ voter out of $N+2$ individuals is, $(N-N^+)/(N+2)$. The $-$ voter has
$N-1$ neighboring voters, but this voter may or may not be connected to the
news sources.  To find the probability that the selected $-$ voter changes
its opinion, we have to consider four possibilities:

 \begin{figure}[ht]
  \centerline{\includegraphics[width=0.975\textwidth]{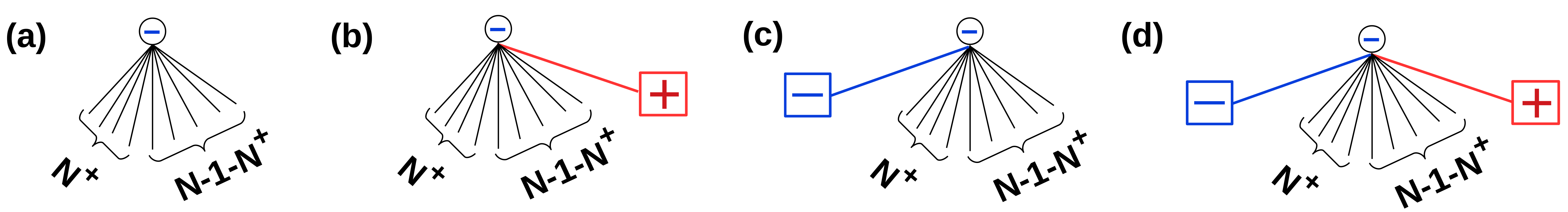}}
  \caption{\small Possibilities for a $-$ voter (circle) to be connected to
    the news sources (squares).  (a) The $-$ voter is connected to $N-1$
    neighboring voters, with $N^+$ having $+$ opinion, but not to any news
    source.  (b) The $-$ voter is connected to a $+$ news source, (c)
    connected to the $-$ news source, and (d) connected to both the news
    sources.}
\label{possibilities}  
\end{figure}

\begin{enumerate}

\item[(a)] The $-$ voter is not connected to any news source
  [Fig.~\ref{possibilities}(a)]: the probability for this configuration is
  $(1-\ell_+)(1-\ell_-)$.  The $-$ voter then adopts opinion from a
  neighboring $+$ voter with probability $N^+/(N-1)$.  The contribution to
  $q^+(x)$ from this configuration is
\begin{subequations}
  \begin{align}
q_1 =(1-\ell_+)(1-\ell_-)\frac{N^+}{N-1}\,.
\end{align}
\item[(b)] The $-$ voter is connected only to the $+$ news source
  (Fig.~\ref{possibilities}(b)), with probability $\ell_+(1-\ell_-)$.  The
  $-$ voter adopts the opinion of a neighboring $+$ voter with probability
  $(1-p)N^+/R$ or adopts the opinion of the $+$ news source with probability
  $p/R$, where $R=p+(N-1)(1-p)$.  Thus the second contribution to $q^+(x)$
\begin{align}
q_2 =\ell_+(1-\ell_-)\frac{(1-p)N^++p}{p+(N-1)(1-p)}\,. \label{q1-cg}
\end{align}

\item[(c)] The $-$ voter is connected only to the $-$ news source
  (Fig.~\ref{possibilities}(c)), with probability $(1-\ell_+)\ell_-$.  The
  probability for the $-$ voter to adopt the opinion of a neighboring $+$
  voter is $(1-p)N^+/R$, where $R=p+(N-1)(1-p)$.  Thus the third contribution
  to $q^+(x)$ is,
\begin{align}
q_3 =(1-\ell_+)\ell_-\frac{(1-p)N^+}{p+(N-1)(1-p)}\,.
\end{align}
\item[(d)] The $-$ voter is connected to both the news sources
  (Fig.~\ref{possibilities}(d)) with probability $\ell_+\ell_-$.  The $-$
  voter thus adopts the opinion of a neighboring $+$ voter with probability
  $(1-p)N^+/R$ or adopts the opinion of the $+$ news source with probability
  $p/R$, where $R=2p+(N-1)(1-p)$. Thus the fourth contribution $q^+(x)$ is
\begin{align}
q_4 =\ell_+\ell_-\frac{(1-p)N^++p}{2p+(N-1)(1-p)} \,.\label{q4-cg}
\end{align}
\end{subequations}
\end{enumerate}

We now write $q^+(x)=N^-(q_1+q_2+q_3+q_4)/(N+2)$ and use $N^-=N-N^+$ ,
$N^+=Nx$, to obtain, after some rearrangement of terms
\begin{subequations}
  \label{qpqm}
\begin{align}
 q^+(x)=\frac{1}{N+2}\left[ANx(1-x) + 2B_+(1-x)\right]\,, \label{qp2}
\end{align}
where $A$ and $B_{+}$ are defined in Eqs.~\eqref{AB}.  Using $+-$ symmetry,
we can also write
\begin{align}
 q^-(x)=\frac{1}{N+2}\left[ANx(1-x) + 2B_-x\right] \,.\label{qm2}
\end{align}
\end{subequations}
In Eqs.~\eqref{qpqm}, the first term inside the bracket accounts for voters
that adopt the opinion of a neighboring voter and the second term accounts
for voters that adopt the opinion of a news source.  Now using
$\delta t= 2/(N+2)$ and Eqs.~\eqref{qpqm} in the definition of
$r^{\pm}(x)=q^{\pm}(x)/\delta t$, we obtain the rates given in
Eq.~\eqref{eqrates}.

\subsection{Voters on the two-clique graph  influenced by two news sources}
\label{2C}

In the two-clique graph, each clique contains $N$ voters, with a $+$ news
source that influences voters on $C_+$ and a $-$ news source that influences
voters on $C_-$.  The entire system thus consists of $2(N+1)$ agents and the
elemental time step is $\delta t= 1/(N+1)$.

Let $x_1$ and $x_2$ be the instantaneous fraction of $+$ voters on $C_+$ and
$C_-$ respectively.  In an elemental time step, define $q_1^{\pm}(x_1,x_2)$
as the probabilities for $x_1$ to change by $\pm \delta x$; similarly,
$q_2^{\pm}(x_1,x_2)$ is the probability for $x_2$ to change by
$\pm \delta x$.  We first evaluate the probability for $x_1$ to increase by
$\delta x$.  For this to occur, a $-$ voter in $C_+$ must be selected which
adopts the opinion of a $+$ neighbor.  The probability to pick a $-$ voter in
$C_+$ out of $2(N+1)$ total agents is $N_1^-/[2(N+1)]$ where $N_1^-=N-N_1^+$
and $N_1^+=Nx_1$ is the number of $+$ voters on $C_+$.  The configurations
that contribute to the probability for the $-$ voter to change its opinion
are:
\begin{enumerate}
\item[(a)] The selected $-$ voter is not connected to the $+$ news source
  (with probability $(1-\ell)$).  This voter has $N+\ell_0-1$ neighboring
  voters, including $N-1$ in the same clique and $\ell_0$ in the other
  clique.  The $-$ voter therefore adopts opinion from a neighboring $+$
  voter in $C_+$ with probability $N_1^+/(N+\ell_0-1)$ or from a $+$ voter in
  $C_-$ with probability $N_2^+\ell_0 /[N(N+\ell_0-1)]$, where $N_2^+=Nx_2$
  is the number of $+$ voters on $C_-$. Thus the first contribution to
  $q_1^+(x_1,x_2)$ is
\begin{subequations}
  \begin{align}
q_1 =(1-\ell)\,\frac{N_1^++\ell_0 (N_2^+/N)}{N+\ell_0-1}\,. \label{q1-tc}
\end{align}

\item[(b)] The selected $-$ voter is connected to the $+$ news source (with
  probability $\ell$).  This $-$ voter has $N+\ell_0$ neighbors, including
  $N+\ell_0-1$ voters and the news source.  The $-$ voter adopts the opinion
  of a neighboring $+$ voter in $C_+$ with probability $(1-p)N_1^+/R$ or the
  opinion of a neighboring $+$ voter in $C_-$ with probability
  $(1-p)\ell_0 N_2^+/(NR)$, where $R=p+(N+\ell_0-1)(1-p)$.  The $-$ voter may
  also adopt opinion from the $+$ news source with probability $p/R$.  Thus
  the second contribution to $q_1^+(x_1,x_2)$ is,
\begin{align}
q_2 =\ell \,\frac{(1-p)[N_1^++\ell_0 (N_2^+/N)]+p}{p+(N+\ell_0-1)(1-p)} \,. \label{q2-tc}
\end{align}
\end{subequations}
\end{enumerate}

We can now write $q_1^+(x_1,x_2)=(N-N_1^+)(q_1+q_2)/[2(N+1)]$.  Using
$N_i^+=Nx_i$ and after some rearrangement of terms, we obtain
\begin{subequations}
  \label{qp1}
\begin{align}
 q_1^+(x_1,x_2)=&\frac{A}{(N+1)}\left[\frac{Nx_1(1-x_1)}{2}+ \frac{\ell_0 (1-x_1)x_2}{2}\right] 
+ \frac{B}{N+1}(1-x_1)\,, \label{qp1-tc}
\end{align}
where $A$ and $B$ are defined in Eqs.~\eqref{coefficients-tc}. In
Eq.~\eqref{qp1-tc}, the term in the square bracket accounts for voters that
adopt the opinion of a neighboring voter.  Inside the square bracket, the
first term accounts for intraclique opinion adoption and the second term
accounts for interclique opinion adoption. The last term in
Eq.~\eqref{qp1-tc} accounts for voters that adopt the opinion from the $+$
news source.

Similarly, we now evaluate the probability for $x_1$ to decrease by
$\delta x$.  For this to occur, a $+$ voter in $C_+$ has to adopt the opinion
of a $-$ neighbor.  Because the $-$ news source influences voters only in
$C_-$, a $+$ voter in $C_+$ can change opinion by either adopting the opinion
from a neighboring $-$ voter in either $C_+$ or $C_-$.  Following the steps
that led to Eq.~\eqref{qp1-tc}, we find
\begin{align}
 q_1^-(x_1,x_2)=\frac{A}{N+1}\left[\frac{Nx_1(1-x_1)}{2}+ \frac{\ell_0 x_1(1-x_2)}{2}\right]\,.   \label{qp2-tc}
\end{align}
\end{subequations}
We use symmetry to find the probability for $x_2$ to change by
$\pm \delta x$.  Their explicit forms are,
\begin{subequations}
  \label{qp4}
  \begin{align}
 q_2^+(x_1,x_2)=&\frac{A}{N+1}\left[\frac{Nx_2(1-x_2)}{2}+ \frac{\ell_0x_1(1-x_2)}{2}\right]  \,,\label{qp3-tc}\\
  q_2^-(x_1,x_2)=&\frac{A}{N+1}\left[\frac{Nx_2(1-x_2)}{2}+ \frac{\ell_0(1-x_1)x_2}{2}\right] 
  +\frac{B}{N+1}\,x_2   \label{qp4-tc}
\end{align}
\end{subequations}

We use $q_i^{\pm}(x_1,x_2)$ in Eqs.~\eqref{qp1} and \eqref{qp4}, and the time
step $\delta t=1/(N+1)$ to determine the opinion change rates. Similar to
Eq.~\eqref{rate1}, we define the rate for $x_i$ to change by $\pm \delta x$
as $r_i^{\pm}(x_1,x_2)=q_i^{\pm}(x_1,x_2)/\delta t$.  Using this definition,
together with the probabilities $q_i^{\pm}(x_1,x_2)$ and the time step
$\delta t$, we obtain Eqs.~\eqref{rates-tc}.

\section{ Polarization time in the complete graph}
\label{conditional-polarization-vm}

To compute the polarization time for the complete graph, Eq.~\eqref{Tp2}, we
need the quantity $\tau$ in this equation.  In turn, $\tau$ is just the
conditional polarization time in the VM.  When the initial fraction of $+$
voters is $y$, with $0<y<1/2$, we first define the conditional polarization
probability to reach $x=\frac{1}{2}$ without hitting $x=0$ as
$E_{\frac{1}{2}}(y)$.  Similarly, the conditional time to reach the polarized
state $x=\frac{1}{2}$ without hitting $x=0$ is $T_{\frac{1}{2}}(y)$.

The conditional probability satisfies the backward equation Eq.~\eqref{Ep}
subject to the boundary conditions $E_{\frac{1}{2}}(0)=0$ and
$E_{\frac{1}{2}}(\frac{1}{2})=1$.  Substituting the drift velocity $V(x)=0$
and diffusion coefficient $D(x)=x(1-x)/2N$ of the VM into Eq.~\eqref{Ep}, we
obtain $E_{\frac{1}{2}}(y)=2y$.  Similarly, the product
$T(y)\equiv T_{\frac{1}{2}}(y)E_{\frac{1}{2}}(y)$, satisfies the backward
equation~\cite{R01},
\begin{align}
T(y) = \varepsilon T(y+\delta y)+(1-\varepsilon) T(y-\delta y)+E_{\frac{1}{2}}(y)dt \,.\label{backward-cond}
\end{align} 
Here $\varepsilon$ is the probability for $y$ to increase by $\delta y$ and
the transition time to leave the state $y$ is $dt$, as given in
Sec.~\ref{formalism}.  Expanding Eq.~\eqref{backward-cond} in a Taylor series
to second order in $\delta y=1/N$ gives
\begin{align}
 V(y) \frac{\partial T(y)}{\partial y}+ D(y) \frac{\partial^2 T(y)}{\partial y^2}=-E_{\frac{1}{2}}(y)\,. \label{ldagger-A}
\end{align}
Solving Eq.~\eqref{ldagger-A} subject to the boundary conditions
$T(0)=T(\frac{1}{2})=0$  gives
\begin{align}
  T_{\frac{1}{2}}(y)= T(y)/E_{\frac{1}{2}}(y)=
  -2N\left[\frac{1-y}{y} \ln (1-y)+\ln 2\right]\,.
\end{align}
For the initial condition $y=\frac{1}{N}$, we have
$\tau=T_{\frac{1}{2}}\big(\frac{1}{N}\big)\approx 2N(1-\ln 2)$ and
$E_{\frac{1}{2}}\big(\frac{1}{N}\big)=\frac{2}{N}$.  These results give the
polarization time in Eq.~\eqref{Tp2}.

\section{Characteristic times on the two-clique graph}
\label{conditional-consensus}

\subsection{Consensus time}

To compute the conensus time for the two-clique graph,
Eq.~\eqref{tc-twoclique}, we need the quantity
$\tau\equiv T_-\big(1-\frac{1}{N}\big)$ in this equation.  Here $T_-(y)$ is
the conditional time for a population on the complete graph that is
additionally influenced by a single $+$ news source to first reach $-$
consensus, without previously reaching $+$ consensus, when the initial
fraction of $+$ voters is $y$ (and vice versa for $T_+(y)$).  Following the
same steps that led in Eq.~\eqref{ldagger-A}, the product
$E_{\pm}(y)T_{\pm}(y)$ satisfies
\begin{align}
 \label{ldagger-B}
  V(y) \frac{\partial(E_{\pm}(y) T_{\pm}(y))}{\partial y}
  + D(y) \frac{\partial^2 (E_{\pm}(y) T_{\pm}(y))}{\partial y^2}=-E_{\pm}(y)\,,
\end{align}
subject to the boundary conditions
$E_{\pm}(0) T_{\pm}(0)=E_{\pm}(1) T_{\pm}(1)=0$.  For the complete graph with
a single news source, $V(x)$ and $D(x)$ are given by Eq.~\eqref{VD32}, from
which we obtain
\begin{align}
 T_-(y)=\begin{cases}
 -2N\left[{\rm Li}_{2}(y)-{\rm Li}_{2}(a)\right]+4N\left[\frac{{\rm Li}_{3}(y)-{\rm Li}_{3}(1)}{\ln y}+\frac{{\rm Li}_{3}(1)-{\rm Li}_{3}(a)}{\ln a}\right] ~&~ \alpha=1,\\ ~\\
 -2N\left[\frac{y}{1-y}\ln y -\frac{\ln a}{N-1}\right]        ~  &~ \alpha=2.
       \end{cases}
       \label{conditional-minus-single}
\end{align}
where $a=\alpha/2N$, and $L_2(y)$ and $L_3(y)$ are the dilogarithm and
trilogarithm functions respectively~\cite{dilog,trilog}.

For the initial condition $y=1-\frac{1}{N}$, $\tau \approx (\pi^2/3) N $ for
$\alpha=1$ and $\tau \approx 2 N $ for $\alpha=2$.  Thus $\tau$ grows
linearly with $N$ for both $\alpha$ values and is subdominant in
Eq.~\eqref{tc-twoclique}.  To show that $\tau$ is subdominant for large
$\alpha$, we will make use of the identity
\begin{align}
T_{\rm con}(y)=E_-(y)T_-(y)+E_+(y)T_+(y)\,.
\label{identity}
\end{align}
We first find a heuristic upper bound for $T_{\rm con}(1-\frac{1}{N})$ and
then use this to find an upper bound on $\tau$.  Since the clique is
influenced by a single news source, the effective potential
\eqref{phi-single} monotonically drives the opinion state $x$ towards 1.
From Eq.~\eqref{tc-amass}, $T_{\rm con}(1-\frac{1}{N})$ is a decreasing
function of $\alpha$ and is of the order of 1 when $\alpha=2$.  We argue that
this decrease continues for larger $\alpha$.  Indeed, for a uniformly biased
random walk on the interval $[0,1]$ that starts near $x=1$, it is known that
the time to reach the boundary at $x=1$ decreases as the bias
increases~\cite{R01}.  Using the hypothesis that $T_{\rm con}(1-\frac{1}{N})$
continues to decrease as $\alpha$ increases in \eqref{identity}, we can write
$E_-(1-\frac{1}{N})T_-(1-\frac{1}{N}) \leq \mathcal{O}(1)$.  Now using the
exit probability Eq.~\eqref{exit-amass}, we obtain
$\tau \leq \mathcal{O}(N^{\alpha})$ for $\alpha>2$.  Consequently, $\tau$
makes a subdominant contribution to the consensus time in
Eq.~\eqref{tc-twoclique} for large $\alpha$.

\subsection{Polarization time}

To compute the polarization time for the two-clique graph,
Eq.~\eqref{Tp2-tc}, we again need the quantity $\tau$ in this equation.  Here
$\tau$ coincides with conditional time $T_+\big(\frac{1}{N}\big)$ in the VM
on the complete graph with no news source.  For this VM, $V(x)=0$,
$D(x)=x(1-x)/2N$, and the exit probability is $E_+(y)=y$.  Using these in
Eq.~\eqref{ldagger-B} now gives
\begin{align}
  T_+(y)&=-2N\frac{(1-y)}{y}\ln (1-y)\,,
\end{align}
so that $\tau \equiv T_+\big(\frac{1}{N}\big)\approx 2N$.

\bigskip

\newcommand{\newblock}{}

\end{document}